%% file: main.tex
\documentclass[sigconf]{acmart}

\input{setup/packages}
\input{setup/custom_commands}

\acmSubmissionID{8660}

\input{defines/defines}

\input{setup/metadata}

\begin{document}

\title[Narrative Keyframing for Generative Creative Writing]{Narrative Keyframing for Generative Creative Writing}

\input{setup/authors}

\begin{abstract}
\input{sections/0-abstract}
\end{abstract}

\input{setup/ccs}


\input{figures/fig-teaser}

\maketitle

\input{sections/1-introduction-final}
\input{sections/3-background}
\input{sections/2-related-work}

\input{sections/4-system-design}
\input{sections/5-technical-evaluation}
\input{sections/6-user-study}
\input{sections/7-discussion}
\input{sections/8-conclusion}

\balance
\bibliographystyle{ACM-Reference-Format}
\bibliography{reference}

\clearpage
\appendix
\input{sections/z-appendix}

\end{document}

%% file: setup/packages.tex
\usepackage{acmart-taps}

\usepackage{tabularray}
\usepackage{multirow} 
\usepackage{array} 
\newcolumntype{C}[1]{>{\centering\arraybackslash}p{#1}} 

\usepackage{float} 

\usepackage{siunitx} 
\usepackage{etoolbox} 
\usepackage{dcolumn} 
\usepackage{booktabs} 

\usepackage{fontawesome5}
\usepackage{svg}
\usepackage{xcolor} 
\usepackage{soul} 
\soulregister{\cite}7 
\soulregister{\citep}7
\soulregister{\citet}7
\soulregister{\ref}7
\soulregister{\pageref}7

\newsavebox{\myfancyboxbox}

\newenvironment{myfancybox}
{%
  \par\addvspace{8pt}%
  \noindent
  \begingroup
  \setlength{\fboxrule}{0.8pt}%
  \setlength{\fboxsep}{8pt}%
  \begin{lrbox}{\myfancyboxbox}%
  \begin{minipage}{%
    \dimexpr\linewidth-2\fboxsep-2\fboxrule\relax
  }%
  \setlength{\parskip}{1pt}%
}
{%
  \end{minipage}%
  \end{lrbox}%
  \fcolorbox{black}{light-bg}{\usebox{\myfancyboxbox}}%
  \endgroup
  \par\addvspace{8pt}%
}

\newcommand{\promptbox}[1]{%
  \begingroup
  \setlength{\fboxsep}{3pt}%
  \colorbox{grey-bg}{#1}%
  \endgroup
}

\usepackage{xspace} 

\usepackage{bm} 
\newrobustcmd*{\bftabnum}{ %
  \bfseries
  \sisetup{output-decimal-marker={\textmd{.}}} %
}

\usepackage{hyperref}
\usepackage{cleveref}

\usepackage{tikz}
\usetikzlibrary{arrows.meta,positioning,fit,backgrounds,calc}

%% file: setup/custom_commands.tex
\definecolor{oxfordblue}{rgb}{0.0, 0.13, 0.28}
\definecolor{harvardcrimson}{rgb}{0.79, 0.0, 0.09}
\definecolor{dartmouthgreen}{rgb}{0.05, 0.5, 0.06}
\definecolor{princetonorange}{rgb}{1.0, 0.56, 0.0}
\definecolor{yaleblue}{rgb}{0.06, 0.3, 0.57}
\definecolor{usccardinal}{rgb}{0.6, 0.0, 0.0}
\definecolor{uclablue}{rgb}{0.33, 0.41, 0.58}
\definecolor{msugreen}{rgb}{0.09, 0.27, 0.23}
\definecolor{cornellred}{rgb}{0.7, 0.11, 0.11}
\definecolor{pomegranate}{RGB}{192, 57, 43}
\definecolor{anti-pomegranate}{RGB}{43,178,192}
\definecolor{alizarin}{RGB}{231, 76, 60}
\definecolor{anti-belize}{RGB}{185, 41, 56}
\definecolor{belize}{RGB}{41, 128, 185}
\definecolor{sky}{RGB}{52, 152, 219}
\definecolor{green}{RGB}{22, 160, 133}
\definecolor{anti-green}{RGB}{160,22,118}
\definecolor{turquoise}{RGB}{26, 188, 156}
\definecolor{pumpkin}{RGB}{211, 84, 0}
\definecolor{anti-pumpkin}{RGB}{0,22,211}
\definecolor{carrot}{RGB}{230, 126, 34}
\definecolor{wisteria}{RGB}{142, 68, 173}
\definecolor{anti-wisteria}{RGB}{99,173,68}
\definecolor{amethyst}{RGB}{155, 89, 182}
\definecolor{nephritis}{RGB}{39, 174, 96}
\definecolor{anti-nephritis}{RGB}{174,39,117}
\definecolor{grey-bg}{RGB}{242,242,235}
\definecolor{light-bg}{RGB}{249,249,249}
\definecolor{extended-blue}{RGB}{59,130,246}
\definecolor{extended-red}{RGB}{239,68,68}
\definecolor{extended-orange}{RGB}{249,115,22}
\definecolor{extended-violet}{RGB}{99,102,241}
\definecolor{extended-green}{RGB}{16,185,129}

\newcommand{\revision}[1]{{#1}}

\newcommand{\eg}{e.g.,\ }

\newcommand{\Tool}{Narrative keyframing\xspace}
\newcommand{\tool}{narrative keyframing\xspace}

\AtBeginDocument{ %
  \providecommand\BibTeX{{ %
    \normalfont B\kern-0.5em{\scshape i\kern-0.25em b}\kern-0.8em\TeX}}}

\newcommand{\namedparagraph}[1]{\vspace{0.2cm}\noindent\textbf{#1:}}

\newcommand{\formatcaption}[2]{\textit{#1} \textmd{#2}}

\newcommand{\fig}[1]{Fig. {#1}}
\newcommand{\figref}[1]{\fig{\ref{#1}}}
\newcommand{\tab}[1]{Table {#1}}
\newcommand{\tabref}[1]{\tab{\ref{#1}}}

\usepackage{enumitem}
\newenvironment{packeditemize}{
\begin{itemize}[leftmargin=0.5cm]
\setlength{\itemsep}{1pt}
\setlength{\parskip}{2pt}
\setlength{\parsep}{0pt}
}{\end{itemize}}

\usepackage{pifont}  



%% file: defines/defines.tex
\newif\ifsubmit
\submitfalse

\input{defines/comments}

\input{defines/commands}

%% file: defines/comments.tex
\ifsubmit
\definecolor{author_colorA}{rgb}{0,0.5,1}
\definecolor{author_colorB}{rgb}{0.2,.64,0}
\definecolor{author_colorC}{rgb}{1,0,1}
\definecolor{author_colorD}{rgb}{0,1,1}
\definecolor{changes_color}{rgb}{0.05,0.5,0.3}
\definecolor{mathbrace_color}{rgb}{0.2,0.5,1.0}
\fi

\ifsubmit
    \newcommand{\abe}[1]{}
    \newcommand{\authorB}[1]{}
    \newcommand{\authorC}[1]{}
    \newcommand{\authorC}[1]{}
    \newenvironment{changes}
      {
      }
      {}
    \newcommand{\URGENT}[1]{}
\else
    \newcommand{\abe}[1]{\textbf{\textcolor{blue}{AD: #1}}}
    \newcommand{\authorB}[1]{\textbf{\textcolor{author_colorB}{CS: #1}}}
    \newcommand{\authorC}[1]{\textsf{\textcolor{author_colorC}{[{\bf AUTHORB}: #1]}}}

    \newcommand{\URGENT}[1]{{\textcolor{red}{URGENT:[#1]}}}
\fi

%% file: defines/commands.tex
\newcommand{\newterm}[1]{\emph{#1}}

\newcommand{\beq}{\begin{equation}}
\newcommand{\eeq}{\end{equation}}

%% file: setup/metadata.tex
\copyrightyear{2026}
\acmYear{2026}
\setcopyright{cc}
\setcctype{by}
\acmConference[UIST '26]{The 39th Annual ACM Symposium on User Interface Software and Technology}{November 02--05, 2026}{Detroit, MI, USA}
\acmBooktitle{The 39th Annual ACM Symposium on User Interface Software and Technology (UIST '26), November 02--05, 2026, Detroit, MI, USA}
\acmDOI{10.1145/3830398.3830586}
\acmISBN{979-8-4007-2856-3/2026/11}

%% file: setup/authors.tex

\author{Chao Zhang}
\email{cz468@cornell.edu}
\orcid{0000-0003-4286-8468}
\affiliation{ %
 \institution{Cornell University}
 \city{Ithaca, NY}
 \country{USA}}
 
\author{Abe Davis}
\email{abedavis@cornell.edu}
\orcid{0000-0003-1469-2696}
\affiliation{ %
 \institution{Cornell University}
 \city{Ithaca, NY}
 \country{USA}}

\renewcommand{\shortauthors}{Zhang and Davis}

%% file: sections/0-abstract.tex
%



\revision{
We introduce \emph{narrative keyframing}, an interaction technique for AI-assisted creative writing that lets writers specify different types of narrative constraints at selected moments in a story, then use AI to generate intervening prose.
Inspired by the use of keyframing in animation, narrative keyframing offers a flexible way to connect story planning with adaptive control over generated text. We explore three types of keyframes: \emph{plot keyframes} define significant events in a story, \emph{character keyframes} represent how individual characters change over the narrative, and \emph{perspective keyframes} capture how individual characters experience different events through first-person narratives. Plot and character keyframes offer a flexible way to adapt the type of high-level conditioning explored in previous AI writing tools to more customizable, iterative, and fine-scale control, while perspective keyframes add a new way to control characterization and focalization by using first-person narratives as an intermediary.
Through a user study, we show that narrative keyframing supports a more controllable, transparent, and engaging way to use generative AI in creative writing.
}

%% file: setup/ccs.tex
\ccsdesc[500]{Human-centered computing~Interactive systems and tools}

%% file: figures/fig-teaser.tex
\begin{teaserfigure}
 \includegraphics[width=\linewidth]{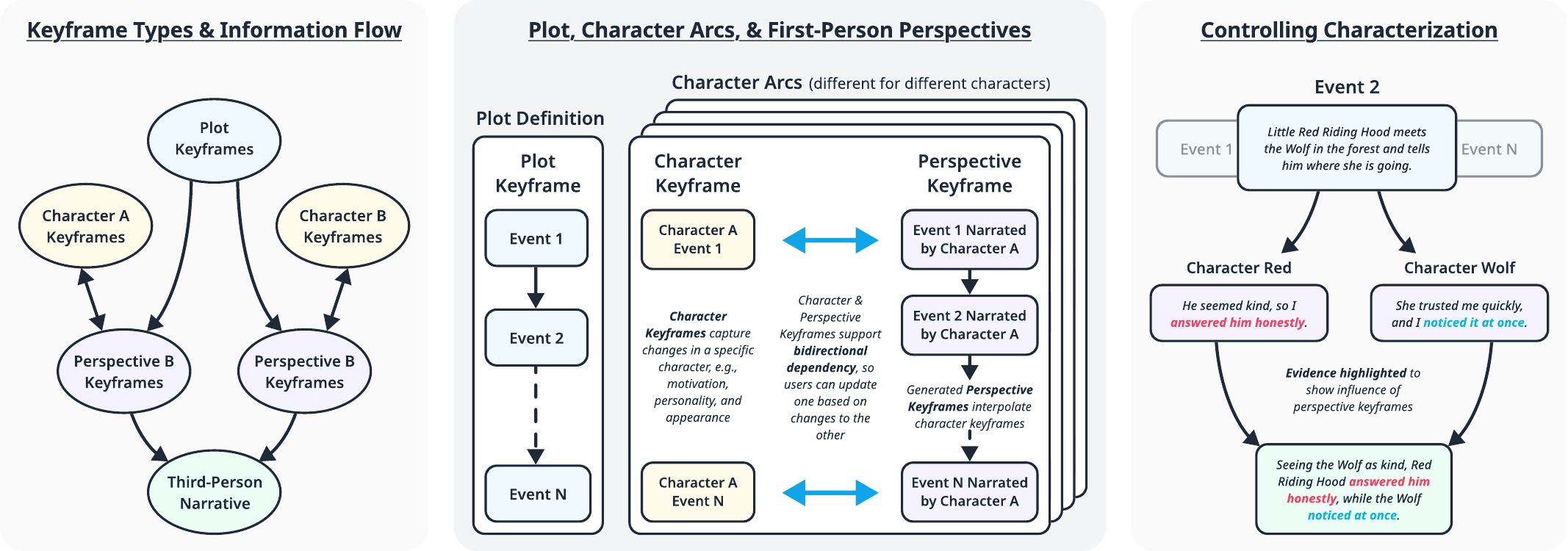}
 \caption{\formatcaption{Narrative Keyframing for Generative Creative Writing.}{Our approach introduces \textit{narrative keyframes} as high-level intermediate representations that connect story planning to narrative generation through three linked forms: plot keyframes, character keyframes, and perspective keyframes (Left). Writers define plot keyframes across events, specify character keyframes to capture how each character changes, and generate first-person perspective keyframes that can be iteratively refined in relation to character states (Center). During story generation, selected evidence from each character's perspective keyframes provides traceability for how characterization decisions shape the resulting third-person narrative (Right).}} 
 \Description{A three-panel schematic illustrates the Narrative Keyframing workflow for generative creative writing. The left panel, titled “Keyframe Types & Information Flow,” shows plot keyframes at the top feeding into two branches of character-specific keyframes and perspective keyframes, which then converge into a final third-person narrative. The middle panel, titled “Plot, Character Arcs, & First-Person Perspectives,” expands this process: a sequence of plot keyframes defines events from Event 1 to Event N; for each event, character keyframes represent how a character changes over time, and corresponding perspective keyframes present first-person narrations of those events. Two-way arrows between character and perspective keyframes indicate that they can be iteratively revised in relation to one another. The right panel, titled “Controlling Characterization,” gives an example for Event 2 from Little Red Riding Hood: the plot event states that Red meets the Wolf in the forest and tells him where she is going. Separate perspective keyframes for Red and the Wolf contain first-person interpretations of the event, with selected phrases highlighted. These highlighted phrases are then carried into a generated third-person sentence below, showing how evidence from perspective keyframes shapes the final narrative portrayal of both characters.}
 \label{fig:teaser_figure}
\end{teaserfigure}

%% file: sections/1-introduction-final.tex
\section{Introduction}

\revision{
In animation, the practice of keyframing offers an incredibly flexible and efficient way to balance automation with creative control. The basic idea is simple: most of the important details in an animation can be derived from constraints on a sparse set of key moments, from which the rest of the animation can be interpolated~\cite{Lasseter87}. 
By controlling the type and distribution of such constraints across a timeline, users can focus creative effort where it is most necessary and leverage automation where it is most appropriate.\looseness=-1

Analogously, creative writing often follows a similar workflow, with writers planning out properties of key moments in a story before connecting those moments with actual prose. In literary theory, this can be described as developing \newterm{plot}, which encompasses the core events of a story, before writing \newterm{narrative}, which encompasses how those events are ultimately presented to the reader~\cite{balNarratologyIntroductionTheory2004,genetteNarrativeDiscourseEssay1990,forsterAspectsNovel1927}. 

Recent AI writing systems have leveraged high-level story descriptions, character specifications, or plot outlines to condition the generation of narrative prose (\eg \cite{qinCharacterMeetSupportingCreative2024,schmittCharacterChatSupportingCreation2021,mirowskiCoWritingScreenplaysTheatre2023}). However, these systems typically treat such conditioning as a set of static prompts or global conditions, limiting fine-grained control over individual story elements and how they change throughout the generated narrative.\looseness=-1

Inspired by animation keyframing (\figref{fig:analogy}), we introduce \emph{narrative keyframing}, a new interaction technique for AI-assisted writing. 
Just as keyframes in animation let artists specify key changes to different animation properties across a timeline, we introduce \emph{narrative keyframes} as a way for authors to specify key changes to different narrative properties across a story.
We explore three types of narrative keyframes: \emph{plot keyframes}, \emph{character keyframes}, and \emph{perspective keyframes}.
Plot keyframes represent events that take place in a story, while character keyframes represent how individual characters change across the narrative. Together, these two types of keyframes generalize the global plot and character conditioning explored in prior work by providing more adaptive and fine-grain event-level control over generated text.
Our third type of keyframes, perspective keyframes, uses written or generated first-person prose to capture how different characters experience particular events.
Perspective keyframes introduce first-person narratives as a novel intermediate representation for controlling characterization and focalization in subsequently generated prose. 
Authors can select and recombine elements from different perspective keyframes that represent how different characters experience a common event to balance how that event is portrayed in a third-person narrative.\looseness=-1

We evaluate our approach through a technical assessment that combines automatic metrics and expert judgments, as well as a user study with 12 writers. Results show that our system produces stories with higher overall quality and richer characterization than a standard LLM baseline, and supports a more controllable, transparent, and engaging writing experience.

\input{figures/fig-analogy}
}

%% file: figures/fig-analogy.tex
\begin{figure}
  \centering
  \includegraphics[width=\linewidth]{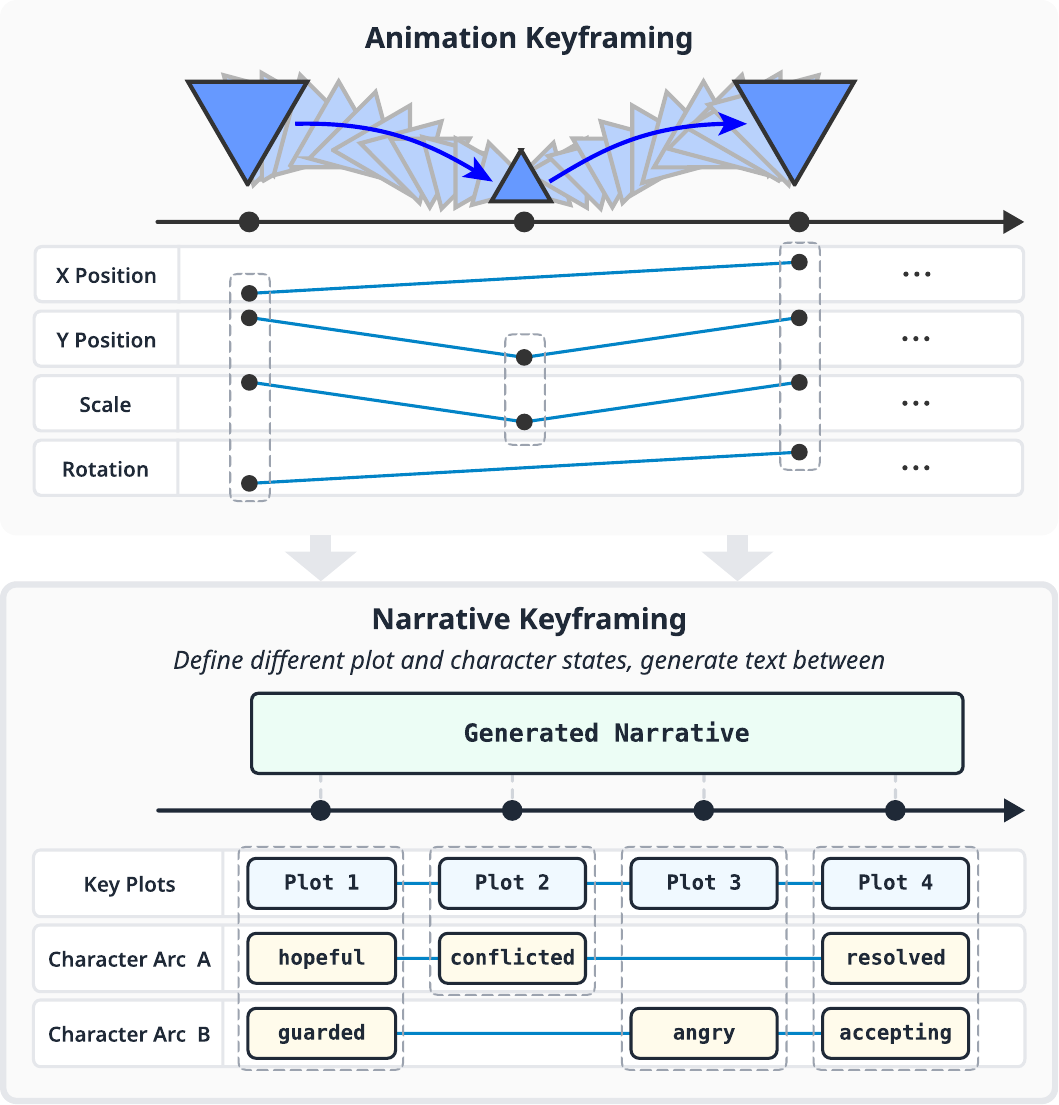}
  \caption{\revision{\formatcaption{From animation keyframing to narrative keyframing.}{Animation keyframing specifies the values of different animation properties at selected points in time and interpolates the states between them. \tool applies the same interaction principle to generative creative writing: users define key plot events and character states at important points in a story, which jointly condition the generation of narrative text between those points.}}}
  \Description{A two-panel diagram illustrating the analogy between animation keyframing and narrative keyframing. The upper panel shows a triangular object moving, rotating, and changing size across three animation keyframes, with lighter intermediate poses generated between them. Four aligned tracks labeled X Position, Y Position, Scale, and Rotation show property values at selected keyframes connected by interpolated lines. Two downward arrows connect this example to the lower panel, titled Narrative Keyframing. The lower panel shows four plot keyframes arranged along a shared timeline and aligned with rows for Key Plots, Character Arc A, and Character Arc B. The plot row contains Plot 1 through Plot 4. Character Arc A develops from hopeful to conflicted and then resolved, while Character Arc B develops from guarded to angry and then accepting. A box labeled Generated Narrative indicates that the plot and character keyframes jointly guide the generation of narrative text between the selected points.}
  \label{fig:analogy}
\end{figure}


%% file: sections/3-background.tex
\input{tables/tab-example}

\revision{\section{Narratological Foundations}
\Tool could, in principle, be applied to many narrative properties. 
In our design, plot events provide the temporal structure on which writers keyframe evolving character states and character-specific perspectives.
We therefore draw on theories of character arcs, characterization, and focalization to motivate these representations and the relationships among them.}

\namedparagraph{Character Arcs}
Character development unfolds throughout a story as various aspects of a character are selectively revealed, reinforced, and transformed at different stages.
Forster's distinction between flat and round characters suggests that flat characters remain relatively stable, while round characters tend to become more complex and develop over the course of the narrative~\cite{forsterAspectsNovel1927}.
This pattern of development for one character across the narrative is often described in creative writing practice as a \emph{character arc}~\cite{weilandCreatingCharacterArcs2016}.
For example, a character may appear self-doubting at the beginning, conflicted in the middle, and confident by the end.
Shaping a character arc requires deciding when particular traits become visible, how they are reinforced across scenes, and how they later change.
This informs our design choice to help writers develop character arcs by defining salient character keyframes at key moments in the story.\looseness=-1

\namedparagraph{Characterization}
Writing theory distinguishes between two complementary processes in characterization: \emph{conceptualization}, in which authors form an understanding of a character's attributes, and \emph{exposition}, in which those attributes are expressed through the narrative medium~\cite{varotsiConceptualisationExpositionTheory2019a}.
In conceptualization, Egri's character ``bone structure''~\cite{egriArtDramaticWriting1995} frames characters as multidimensional constructions spanning \emph{physiology} (\eg age and appearance), \emph{psychology} (\eg personality, values, and motivations), and \emph{sociology} (\eg profession, status, and relationships).
However, rich conceptualization alone does not ensure effective characterization. Story quality also depends on how character attributes are conveyed in specific plots~\cite{garveyCharacterizationNarrative1978,harveyCharacterNovel,varotsiConceptualisationExpositionTheory2019a}.
Rather than being stated explicitly, traits are often conveyed indirectly through a range of narrative ``expositors,'' including physical appearance, actions, thoughts, dialogue, setting, and symbolic elements~\cite{rimmon-kenanNarrativeFictionContemporary2003}.
These theories inform our design choice to support both the development of rich character concepts and their translation into concrete narrative details.\looseness=-1

\namedparagraph{Focalization}
Characterization is shaped by the narrative perspective through which character attributes are made available to readers~\cite{balNarratologyIntroductionTheory2004,genetteNarrativeDiscourseEssay1990}.
Narratology describes this in terms of \emph{focalization}~\cite{balNarrationFocalization2004}: narrative discourse may be organized around what a particular character perceives, knows, and feels, or it may shift across multiple characters within a broader narrative frame.
Within this framing, \emph{first-person narration} provides access to a character's subjective experience, making it useful for exploring expository details such as perception, interpretation, emotion, and self-understanding.
\emph{Third-person narration}, by contrast, provides a broader frame for organizing focalization across multiple characters and scenes.
When writing a third-person story, first-person narration can serve as an intermediate step for externalizing character-specific expository material.
This is similar to point-of-view writing exercises~\cite{jodicleghornWritingExerciseSwitching2010} (also describe as ``method writing''~\cite{grapesMethodWritingFirst}) used by practitioners, in which writers explore a scene from a particular character's perspective before integrating those insights into the broader narrative.
This informs our design choice to let writers explore each character's arc through perspective-specific narration, then compose a final story that integrates those viewpoints.\looseness=-1


%% file: tables/tab-example.tex
\definecolor{redhood}{RGB}{180,100,90}
\definecolor{wolf}{RGB}{90,110,180}

\newcommand{\rh}[1]{%
  \textcolor{redhood}{\emph{\textbf{#1}}}%
}
\newcommand{\wf}[1]{%
  \textcolor{wolf}{\emph{\textbf{#1}}}%
}

\begin{table*}[tb!]
\footnotesize
\centering
\caption{\formatcaption{Example of perspective-based recombination.}{From a single story event, writers generate first-person perspectives for different characters, then selectively recombine evidence from those perspectives to produce third-person narratives with different characterization emphases. \rh{Red text} indicates phrases drawn from Red's perspective, and \wf{blue text} indicates phrases drawn from the Wolf's perspective.}}
\label{tab:keyframing_example}

\begin{tabular}{
  p{0.15\linewidth}
  p{0.54\linewidth}
  p{0.23\linewidth}
}
\toprule
\textbf{Stage}
&
\textbf{Narrative Content}
&
\textbf{Characterization Focus}
\\
\midrule

\textbf{Source Plot}
&
Little Red Riding Hood meets the Wolf in the forest and tells him where she is
going.
&
---%
\\

\midrule
\multicolumn{3}{l}{\textbf{First-Person Perspectives}}
\\
\quad First-Person -- Red
&
He \rh{seemed kind}, so \rh{I answered him honestly}.
\rh{I was curious} and \rh{did not sense any danger}.
&
Red's perspective
\\

\quad First-Person -- Wolf
&
She \wf{trusted me quickly}, and \wf{I noticed it at once}.
\wf{I spoke softly} so she would \wf{keep talking}.
&
Wolf's perspective
\\

\midrule
\multicolumn{3}{l}{\textbf{Third-Person Narratives}}
\\
\quad Third-Person -- A
&
Seeing the Wolf as \rh{kind}, Red Riding Hood \rh{answered him honestly},
while the Wolf \wf{noticed at once} that she \wf{trusted him quickly}.
&
Red's \rh{trust}; Wolf's \wf{opportunism}
\\

\quad Third-Person -- B
&
\rh{Curious} and unable to \rh{sense any danger}, Red Riding Hood kept
speaking with the Wolf, while the Wolf \wf{spoke softly} so that
\wf{she would keep talking}.
&
Red's \rh{curiosity}; Wolf's \wf{manipulation}
\\

\quad Third-Person -- C
&
Because the Wolf \rh{seemed kind} and she \rh{did not sense any danger},
Red Riding Hood lowered her guard, while the Wolf
\wf{noticed it at once} and \wf{spoke softly} to cultivate her trust.
&
Red's \rh{misjudgment}; Wolf's \wf{calculation}
\\

\bottomrule
\end{tabular}

\Description{A three-column table illustrates how first-person perspectives can be recombined into different third-person narratives. The columns are Stage, Narrative Content, and Characterization Focus. The first row gives the source plot event: Little Red Riding Hood meets the Wolf in the forest and tells him where she is going. The next section, labeled First-Person Perspectives, contains one row for Red and one for the Wolf. Red's first-person version emphasizes that the Wolf seemed kind, that she answered him honestly, that she was curious, and that she sensed no danger. The Wolf's first-person version emphasizes that Red trusted him quickly, that he noticed this immediately, and that he spoke softly to keep her talking. The final section, labeled Third-Person Narratives, contains three alternative recombinations. Version A combines Red's sense of kindness and honesty with the Wolf's awareness of her trust, producing a characterization focus of Red's trust and the Wolf's opportunism. Version B combines Red's curiosity and lack of danger with the Wolf's soft speech and intention to keep her talking, producing a focus on Red's curiosity and the Wolf's manipulation. Version C combines Red's impression that the Wolf seemed kind and harmless with the Wolf's quick notice and soft speech, producing a focus on Red's misjudgment and the Wolf's calculation. Overall, the table shows how different pieces of evidence drawn from separate first-person perspectives can be selectively recombined to create different third-person portrayals of the same event.}
\end{table*}

%% file: sections/2-related-work.tex
\section{Related Work}
\looseness=-1

\revision{
\subsection{Keyframing in Animation}
\input{sections/animationkeyframes}
}


\revision{
\subsection{Intelligent Writing Interfaces}
The HCI community has long been interested in intelligent writing tools~\cite{leeDesignSpaceIntelligent2024} that support writers across a range of writing tasks, including brainstorming ideas~\cite{geroSparksInspirationScience2022,schmittCharacterChatSupportingCreation2021,chouTaleStreamSupportingStory2023,zhangStoryDrawerChildAI2022}, planning outlines~\cite{wanPolymindParallelVisual2025a,riedlVignettebasedStoryPlanning2008}, drafting content~\cite{zhangWordsWidgetsControllable2026,dhillonShapingHumanAICollaboration2024,hoqueHaLLMarkEffectSupporting2024,jakeschCoWritingOpinionatedLanguage2023,kimAuthorsValuesAttitudes2024,kimMechanicalNovelCrowdsourcing2017,zhangMathemythsLeveragingLarge2024}, and refining text~\cite{zhangSynthiaVisuallyInterpreting2025,itoUseAIpoweredRewriting2023,leeInteractiveChildrenStory2022,rezaABScribeRapidExploration2023,turkayIteroRevisionHistory2018}.
These tools span diverse genres, including argumentative writing~\cite{zhangVISARHumanAIArgumentative2023,zhangFrictionDecipheringWriting2025}, story writing~\cite{chungTaleBrushSketchingStories2022a,yuanWordcraftStoryWriting2022,huangHeteroglossiaInSituStory2020}, and scientific writing~\cite{shenConvXAIDeliveringHeterogeneous2023a,sunMetaWriterExploringPotential2024}.
Within this broader design space, recent work has explored interaction metaphors drawn from established creative and design practices to develop visual interfaces for human-AI co-writing.
Rather than relying solely on natural language prompts, these systems externalize aspects of the writing process into visual representations that allow writers to control AI generation.
Different metaphors emphasize different forms of authorial control, ranging from high-level planning~\cite{chungTaleBrushSketchingStories2022,zhangNarrixRemixingNarrative2026,chungPatchviewLLMpoweredWorldbuilding2024a} to local text revision~\cite{massonTextoshopInteractionsInspired2025,shenTexterialTextasMaterialInteraction2026}.\looseness=-1

For example, TaleBrush~\cite{chungTaleBrushSketchingStories2022} supports control over generated stories by editing 2D curves that represent attributes such as surprise, while CharacterChat~\cite{schmittCharacterChatSupportingCreation2021} and CharacterMeet~\cite{qinCharacterMeetSupportingCreative2024} help writers construct global character personas via role-play. 
These methods operate at a more global level, while ours allows for specification of detailed character and plot conditions at specific plot points. 
Another line of research focuses on local text-level control.
For instance, Texterial~\cite{shenTexterialTextasMaterialInteraction2026} conceptualizes text as clay, allowing users to refine generated content through gestural sculpting, while Textoshop~\cite{massonTextoshopInteractionsInspired2025} borrows interactions from drawing software to support editing operations such as shortening and reordering text. 
These systems provide useful mechanisms for revising surface-level text but do not explicitly support control over the narrative dimensions of a story.\looseness=-1

A group of systems more closely related to our work explores how writing can be represented as the manipulation of visual structures composed of discrete writing elements.
For example, VISAR~\cite{zhangVISARHumanAIArgumentative2023} and Polymind~\cite{wanPolymindParallelVisual2025a} draw inspiration from node-based visual programming, enabling users to control text generation through interconnected nodes. 
However, VISAR focuses on supporting logical structure in argumentative writing, while Polymind emphasizes constructing AI workflows from microtasks such as ``summarize'' and ``brainstorm,'' rather than controlling the progression of a narrative. 
Dramatron is closer to our work in its staged decomposition of story generation into narrative elements, including characters, plot, locations, and dialogue. However, this decomposition is controlled through sequential user prompts in a Colab environment, limiting users to comparatively rigid interactions within a fixed hierarchical decomposition of the story.\looseness=-1

Uniquely, our keyframing interaction allows authors to control different types of narrative constraints (\eg plots and characters) at selected moments in a story while leaving the progression between them to AI generation. This approach provides fine-grained yet flexible control over character arcs and narrative development throughout the writing process.\looseness=-1
}

\subsection{Interactive Characterization Tools}
Recent HCI systems have explored characterization as an interactive process, often by enabling writers to build characters through simulation and dialogue with LLM-powered personas. For example, CharacterChat~\cite{schmittCharacterChatSupportingCreation2021} supports character creation through conversational roleplay and progressive manifestation, while CharacterMeet~\cite{qinCharacterMeetSupportingCreative2024} extends this idea to support writers throughout the broader process of story character construction via chatbot avatars. 
Related systems similarly use persona-driven or multi-agent character simulation to help writers explore character traits, backstory, and possible narrative developments~\cite{fuYourStoryCoCreative2025,parkConstellaSupportingStorywriters2026,wangStoryVerseCoauthoringDynamic2024,cavazzaCharactersSearchAuthor2001}.
These systems primarily help users define global, static character sheets, but offer limited support for understanding and controlling how an established character profile is expressed in a story or how character traits evolve across plot events. 
\revision{To address this gap, our work uses first-person perspectives as an intermediate representation for conditioning narrative generation. 
First-person perspectives instantiate abstract character definitions into concrete narrative text, allowing users preview how a character is portrayed before selecting what to emphasize in the final prose. 
First-person perspectives also fit naturally within our keyframing interaction by enabling users to shape the progression of a character arc at key plot events.
To our knowledge, ours is the first work to explore first-person character narratives as an intermediate representation for controlling AI-assisted creative writing.}

%% file: sections/animationkeyframes.tex
The practice of animation keyframing dates back long before the invention of computers. The earliest version was used as a pre-visualization strategy in which artists would draw keyframes to plan the flow of an animation before committing effort to drawing all the remaining frames. When animators started working in teams, this became a way to divide labor: a lead artist would draw detailed keyframes, which assistants or trainees would then interpolate~\cite{Lasseter87}. With the advent of computers, the practice of keyframing evolved into a general technique for interpolating between artist-specified constraints across an animation timeline.
Modern tools let users create different types of keyframes to control different properties of an animation (\eg position, rotation, color), and users can adjust the density of keyframes over a timeline to adaptively balance interpolation with finer-grained creative control.

Abstractly, we can think of keyframing as a flexible and efficient way to balance automation with creative control across a timeline. It requires only two things: a way to localize creative constraints on the timeline and a way to interpolate between those constraints. 
The key insight of our work is that recent advances in generative language models make an analogous form of interpolation possible for narrative. 
Language models can generate coherent narrative developments and prose between sparse constraints specified at important points in a story.
This allows writers to anchor major plot events and character states while delegating intermediate developments to AI generation. 
As in animation, writers can vary both the type and density of keyframes to determine where direct authorial control is most important and where greater automation is appropriate. 

%% file: sections/4-system-design.tex
\section{Design Goals}
\revision{Drawing on the narratological theories and related HCI work discussed above, we derive three design goals for our instantiation of \tool.
We use characterization as a concrete design context for exploring how narrative keyframes can support control across story planning, perspective exploration, and narrative generation. The resulting goals concern how writers shape character development across a story, translate abstract character ideas into narrative form, and understand how those decisions are reflected in generated text.}

\begin{packeditemize}
    \item \textbf{[DG1] \textit{Shaping character development across plot points.}} Characterization unfolds over narrative progression rather than appearing all at once~\cite{forsterAspectsNovel1927}. Writers should be able to represent how a character changes at key moments in the story, inspect that progression, and revise it as the narrative develops. This goal follows from theories of character arcs~\cite{weilandCreatingCharacterArcs2016} and motivates support for planning characterization across plot points rather than specifying characters only once at the beginning.
    \item \textbf{[DG2] \textit{Manifesting abstract character traits into concrete narrative evidence.}} Characterization depends not only on defining who a character is, but also on expressing those qualities through narrative details~\cite{garveyCharacterizationNarrative1978,harveyCharacterNovel,varotsiConceptualisationExpositionTheory2019a}. Writers should be able to explore how abstract traits may be realized in language through individual characters' perspectives, then inspect concrete evidence such as actions, thoughts, dialogue, appearance, and setting. This goal follows from theories of conceptualization, exposition~\cite{varotsiConceptualisationExpositionTheory2019a}, and focalization~\cite{balNarrationFocalization2004}, and motivates support for using perspective-specific narration as an intermediate step between character planning and story generation.
    \item \textbf{[DG3] \textit{Tracing how characterization decisions propagate into generated story text.}} Because characterization may be developed through multiple intermediate steps (\eg conceptualization of characters, first-person perspectives), writers should be able to inspect how earlier decisions influence later generations. This includes tracing how character traits and perspective-specific details are carried into the third-person narrative. This goal follows from theories of focalization and perspective~\cite{balNarrationFocalization2004,balNarratologyIntroductionTheory2004,genetteNarrativeDiscourseEssay1990}, and motivates interfaces that make the relationship between characterization inputs and generated story text visible.\looseness=-1
\end{packeditemize}

\section{System Design}

\input{figures/fig-track}
\input{figures/fig-table}
\input{figures/fig-canvas}

Informed by these design goals, we instantiate \tool in an interactive system that connects story planning, character development, perspective exploration, and narrative generation. The system is organized around three linked keyframe types---plot, character, and perspective---that allow writers to specify narrative constraints at key plot points and use them to guide the generation of the final story.
In this section, we describe the overall workflow of this system, the three types of keyframes, and the implementation details.\looseness=-1

\subsection{Workflows and Views}
Our system supports a flexible and iterative workflow that moves from story planning to character and perspective exploration, and finally to narrative generation.
Users (1) create a story outline organized by plot keyframes; (2) define character keyframes to represent how each character develops across the story; (3) generate perspective keyframes to explore how those traits may be expressed in narrative form; (4) select the character traits and textual evidence they want to emphasize; and (5) generate third-person narratives conditioned on those keyframes and selections.
The following subsections describe the features that support each stage of this workflow.

To support this workflow, the interface consists of three coordinated views:
\begin{packeditemize}
    \item The \textbf{Track} view (\figref{fig:track}) supports the main end-to-end workflow, allowing writers to move from outline creation to character keyframes, first-person perspective exploration, and final third-person story generation.
    \item The \textbf{Table} view (\figref{fig:table}) presents the outline, each character's first-person perspective, and the generated third-person narrative side by side for each plot, helping writers compare them in parallel and trace how outline content is developed into the final narrative, as well as how details from perspectives are transformed and incorporated into the third-person narrative.
    \item The \textbf{Canvas} view (\figref{fig:canvas}) supports broader exploration by allowing writers to create multiple character arcs for a single character and combine different arcs across characters to generate alternative versions of the final narrative.
\end{packeditemize}

\subsection{Plot Keyframes}
Our system begins with \emph{plot keyframes} (\figref{fig:track}A), which represent the major events in a story outline. Each plot keyframe acts as a structural anchor for the corresponding character keyframes, perspective keyframes, and generated narrative. This event-based representation helps writers break a story into manageable units, plan character development across events (\textbf{DG1}), and maintain alignment between high-level plot structure and later generated text.\looseness=-1

\subsection{Character Keyframes}
To help writers control character development, our system lets them define \emph{character keyframes} (\figref{fig:track}B) at key plots in the story outline. Each keyframe captures the character's state at a particular moment in the narrative and serves as a building block for shaping that character's arc over the course of the story (\textbf{DG1}).

\namedparagraph{Defining Character Traits}
Based on Egri's character ``bone structure''~\cite{egriArtDramaticWriting1995} and prior work~\cite{schmittCharacterChatSupportingCreation2021}, each keyframe organizes traits into three dimensions: \emph{physiology} (\eg age and appearance), \emph{psychology} (\eg personality, values, and ambitions), and \emph{sociology} (\eg profession, status, and relationships).
For each dimension, users can add their own traits via the plus icon or ask the AI via the sparkles icon to suggest three additional traits based on the existing traits and story outline (\figref{fig:track}C).

\namedparagraph{Interpolating Character Development}
Because writers may not want to manually specify every intermediate character state, the system can automatically interpolate character keyframes (\figref{fig:track}D) between existing ones to suggest how a character may transition over story progression (\textbf{DG1}). These generated keyframes remain fully editable, allowing users to revise, add, or remove traits as needed.\looseness=-1

\subsection{Perspective Keyframes}
To help writers explore how a character's traits may be expressed in narrative form, our system generates first-person \emph{perspective keyframes} (\figref{fig:track}E) from character keyframes. These perspectives externalize character attributes into textual evidences that can later be used to guide the generation of third-person narratives (\textbf{DG2}).

\namedparagraph{Generating First-Person Perspectives}
After creating character keyframes for a character, users can click the play icon to generate first-person perspectives for that character (\figref{fig:track}F).
To guide generation, we incorporate Rimmon-Kenan's typology~\cite{rimmon-kenanNarrativeFictionContemporary2003} of textual indicators of character traits, including \textit{direct definition}, \textit{actions}, \textit{speech}, \textit{appearance}, and \textit{environment}, into the prompt to guide the model to manifest the user-defined character attributes for the corresponding perspective keyframe (\textbf{DG2}).
For example, if a user specifies the trait ``self-doubting,'' the generated perspective may express it through evidence such as hesitation in action (``\textit{I paused before reaching for the door}''), self-questioning in thought (``\textit{What if I get this wrong again?}''), or uncertainty in speech (``\textit{I'm not sure this is a good idea}'').

\namedparagraph{Inspecting and Selecting Textual Evidence}
Given a generated first-person perspective, users can click the search icon (\figref{fig:track}G) to ask the AI to identify textual evidence for the character traits they defined (\textbf{DG2}), based on Rimmon-Kenan's typology~\cite{rimmon-kenanNarrativeFictionContemporary2003}.
Users can then click individual character traits (\figref{fig:track}F) to highlight or hide the corresponding evidence in the perspective keyframes.
Highlighted evidence (\eg \figref{fig:track}G) is selected by default for use in generating the final third-person narrative; and users can click to manually deselect any highlighted passage.

\namedparagraph{Bidirectional Editing Between Character Keyframes and Perspective Keyframes}
Each character keyframe and its corresponding perspective keyframes are bidirectionally linked within an plot.
Users can click the pencil icon (\figref{fig:track}H) to edit a perspective keyframe, which triggers an update to the corresponding character keyframe (\figref{fig:track}I).
Conversely, editing a character keyframe triggers regeneration of the associated perspective keyframe.

\subsection{Third-Person Narratives}
After exploring characters through perspective keyframes and selecting the traits and textual evidence they want to emphasize, users can click the play icon to generate a third-person narrative (\figref{fig:track}J).
Before generation, the system opens a panel for users to review all selected textual evidence from each character's perspective in each plot.
Once users confirm their selections, the system generates a third-person narrative based on the plots and enriches it with the selected textual evidences of character traits.
In the generated narrative, passages derived from different characters' perspectives are highlighted in different colors.
This color coding helps users trace how character traits are expressed in first-person perspectives and how those materials are later synthesized into the final third-person narrative (\textbf{DG3}).
Lastly, users can click the file icon to populate the generated third-person narrative into a Markdown text editor, where they can further review and refine it.

\subsection{Implementation Notes}
The system is built with the Next.js framework, which supports server-side rendering for API calls, including calls to the OpenAI API for prompting pre-trained GPT models, and to the Firebase APIs for logging user events.
We use React Flow to build the node-based canvas and Slate.js to build the text editor.
We instruct \texttt{GPT-4.1} to suggest character traits, interpolate character keyframes, generate first-person perspectives, extract evidence from perspectives, and generate narratives based on selected traits and evidences.
Sample prompts are provided in Appendix~\ref{appendix:technical_details}.
The source code will be open-sourced upon publication.
\looseness=-1

%% file: figures/fig-track.tex
\begin{figure*}
  \centering
  \includegraphics[width=0.95\linewidth]{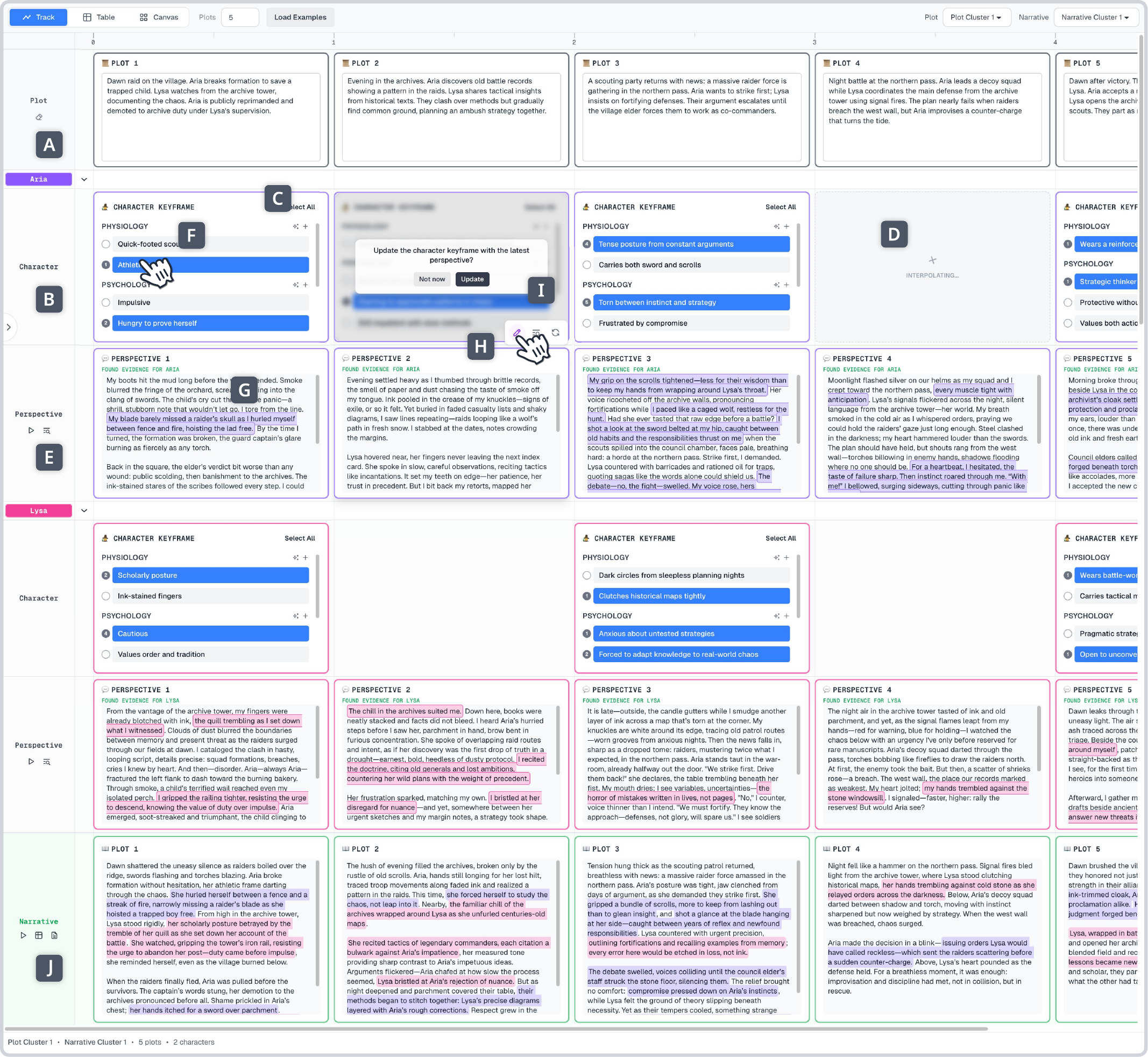}
  \caption{\formatcaption{Track View.}{The Track view supports the main end-to-end workflow from story planning to generation. Writers define (A) plot keyframes, which represent the major events in a story outline. Then, they (B) specify character keyframes for certain plots, and can trigger the AI to (C) suggest new traits or (D) automatically interpolate character development between existing character keyframes. The system generates (E) first-person perspective generates based on plot and character keyframes. Selecting a specific trait (F) highlights its corresponding textual evidence (G) within the perspective keyframe. The system supports bidirectional editing; manually editing a perspective keyframe (H) prompts the system to update the corresponding character keyframe (I). Finally, writers generate a (J) third-person narrative that synthesizes the selected textual evidence from perspective keyframes, with text color-coded by character.}}
  \Description{A screenshot of the Track view shows the system’s main workflow arranged as a grid from left to right across five plot events. The interface is organized into horizontal bands labeled Plot, Character, Perspective, and Narrative, with two characters, Aria and Lysa, shown in separate color-coded sections. In the top row, rectangular plot keyframes summarize major story events in sequence. Beneath them, character keyframe panels list selected traits under categories such as physiology and psychology. Some character panels show manually specified traits, while one column displays an interpolation state, indicating that intermediate character development can be generated automatically between existing keyframes. In the perspective row, first-person perspective keyframes present paragraphs of generated text for each character at each plot event. Within these panels, selected textual spans are highlighted to indicate evidence linked to specific character traits. Clicking a trait in a character keyframe highlights its corresponding evidence in the associated perspective text. Near the center, editing controls and a pop-up dialog illustrate bidirectional updating: after a perspective keyframe is manually edited, the system prompts the writer to update the linked character keyframe. In the bottom row, third-person narrative panels show generated story passages that synthesize selected evidence from the perspective keyframes, with highlighted phrases color-coded by character to indicate their source. Letter annotations A through J mark the major steps of the workflow, from defining plot keyframes, authoring and refining character keyframes, generating and editing first-person perspectives, and producing the final third-person narrative.}
  \label{fig:track}
\end{figure*}

%% file: figures/fig-table.tex
\begin{figure}
  \centering
  \includegraphics[width=\linewidth]{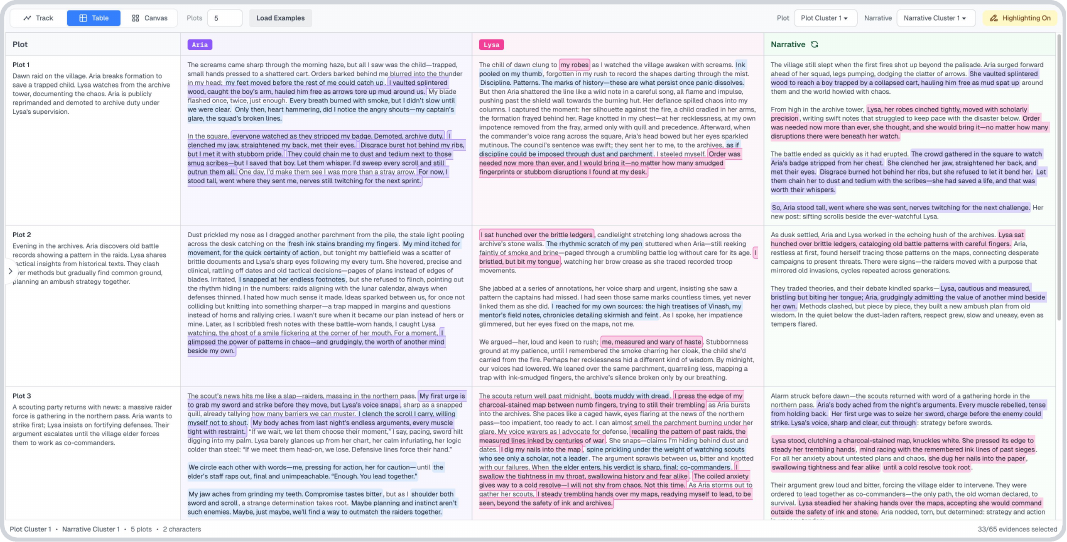}
  \caption{\formatcaption{Table View.}{\revision{The Table View aligns plot keyframes, perspective keyframes, and the generated narrative on a row-by-row basis. Color-coded highlights indicate evidence selected from each character's perspective and show where it is reflected in the final third-person narrative, supporting comparison and traceability across representations.}}}
  \Description{A screenshot of the Table view shows story materials arranged in parallel columns for comparison. The interface is organized into four vertical columns labeled Plot, Aria, Lysa, and Narrative. Each row corresponds to a plot event, beginning with Plot 1 at the top and continuing downward through later events. In the leftmost Plot column, each cell contains a short plot keyframe describing a major event in the story outline. The middle two columns present first-person perspective keyframes for the two characters, Aria and Lysa, shown in purple and pink section headers. These cells contain longer narrative passages written from each character’s perspective. Within the perspective passages, selected phrases are highlighted to indicate evidence chosen for later use. The rightmost Narrative column contains generated third-person story passages that correspond to each plot event. These passages also include color-coded highlighted phrases, showing how evidence from the characters’ first-person perspectives has been incorporated into the final narrative. By aligning plot descriptions, both characters’ perspectives, and the resulting third-person narrative side by side, the view supports close comparison of how plot details and character-specific interpretations are transformed into the final story text.}
  \label{fig:table}
\end{figure}

%% file: figures/fig-canvas.tex
\begin{figure}
  \centering
  \includegraphics[width=\linewidth]{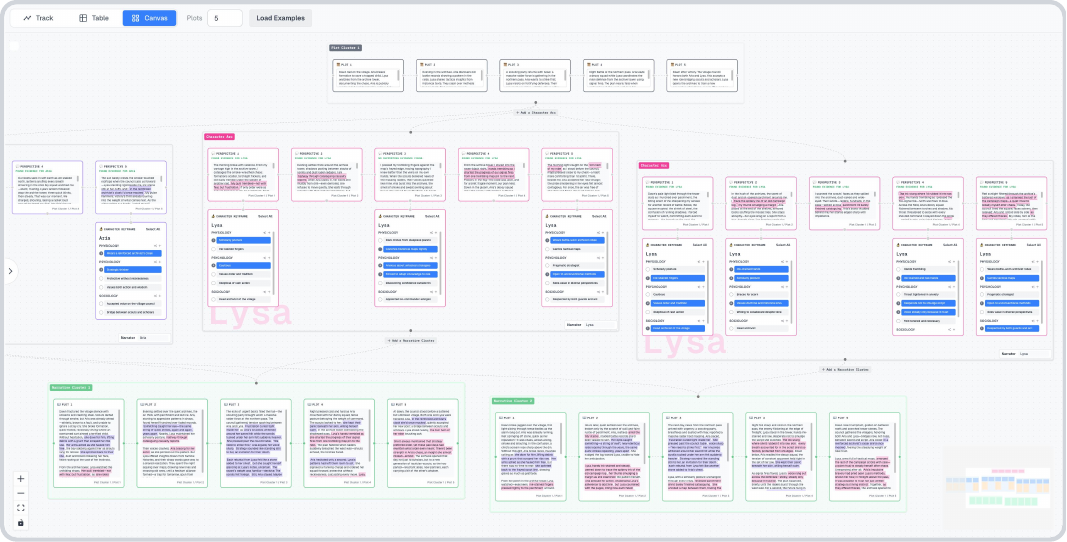}
  \caption{\formatcaption{Canvas View.}{\revision{The Canvas View presents story materials as a node-link graph, making branching character arcs and their recombination explicit. The visualization helps writers compare alternative narrative paths and inspect how different character arc combinations lead to different generated stories.}}}
  \Description{A screenshot of the Canvas view shows a large node-based workspace for exploring alternative story developments. At the top, a row of plot keyframes defines a sequence of story events. Below, the workspace branches into multiple clusters of cards connected by curved lines, indicating different possible paths through the story. In the middle area, several pink clusters represent alternative character arcs for the character Lysa. Each cluster contains multiple first-person perspective keyframes paired with character keyframe cards, allowing different versions of Lysa’s development to be explored across the same sequence of plot events. Small controls between clusters suggest that alternative arcs can be added and connected. At the bottom, green clusters contain generated third-person narrative passages. These narrative clusters combine different character-arc branches to produce alternative story versions. Colored text highlights within the narrative cards indicate evidence drawn from different characters or sources. The overall layout emphasizes branching, recombination, and comparison, showing how writers can create multiple arcs for a single character and then mix different arcs across characters to generate different final narratives.}
  \label{fig:canvas}
\end{figure}

%% file: sections/5-technical-evaluation.tex
\section{System Evaluation}
\revision{Our system instantiates \tool through plot, character, and perspective keyframes, with character development serving as the primary narrative property under writers' control. Accordingly, our evaluation focuses on the system's support for characterization.}
To this end, we evaluated the system through two studies: a technical evaluation (Study 1) of story outputs and a user study (Study 2) of writers' experiences. 
\revision{The goal of Study 1 is to validate that our choice of narratologically-motivated conditioning (traits, first-person perspectives, and evidence-guided recombination) could lead to richer characterization without hurting quality relative to a common outline-to-narrative baseline.
Study 2 served as our primary evaluation of user support, investigating how the system supports characterization in writing practice.}
Together, these studies assess both output quality and the interaction benefits of the system.\looseness=-1

\subsection{Study 1: Technical Evaluation}
Before conducting the user study, we performed a technical evaluation to examine whether our approach could produce higher-quality stories than a vanilla LLM baseline.

\subsubsection{Method}
Here, we describe the method of our technical evaluation, including the materials and metrics.


\namedparagraph{Materials}
To generate stories, we used the ten writing prompts from the \texttt{CoAuthor} dataset~\cite{leeCoAuthorDesigningHumanAI2022a}, supplemented by another ten prompts randomly sampled from the \texttt{WritingPrompts} dataset~\cite{fanHierarchicalNeuralStory2018}.
The full list of prompts is provided in Appendix~\ref{appendix:writing_prompts_tech}.
For each prompt, we first instructed \texttt{GPT-4.1} to generate five diverse outlines featuring two main characters and following a three-act structure~\cite{fieldScreenplayFoundationsScreenwriting2005} (setup, confrontation, and resolution).
We then provided each outline as input to both our system and a vanilla prompt-based LLM baseline, asking each condition to expand the outline into a complete story.
Both conditions used the same underlying model, \texttt{GPT-4.1}.
For our system, we enabled the automatic character trait suggestion feature to generate three traits in each category for each character at each plot.
We then used these traits to generate first-person perspectives and randomly selected two pieces of evidence per character per plot to generate the final third-person story.\looseness=-1

\namedparagraph{Metrics}
We evaluated the quality of the generated stories from the two conditions using the \texttt{WQRM-PRE} model from Chakrabarty et al.~\cite{chakrabartyAISlopAIPolishAligning2025}, which was trained on expert preference data and has been shown to align with expert judgments.
This model produces a scalar quality score for each story.
We used these scores for both pairwise comparisons (between the two stories generated from the same prompt) and a paired t-test over the corpus.

To complement the automatic evaluation, we also conducted a human validation on a randomly sampled subset of 30 story pairs from the two conditions.
Three independent raters with creative-writing experience recruited from Prolific compared the paired stories in randomized order, with condition identities removed.
All three human raters self-reported that they were experienced professional writers.
Two had 4--6 years of experience, and one had 7--10 years, including work in screenwriting and editorial evaluation.
For each pair, raters indicated (1) which story they preferred overall and (2) which story exhibited richer characterization.
These two dimensions were chosen to validate both the general quality signal captured by the automatic evaluator and the characterization-focused contribution of our system.
Each rater was compensated with \$20.
\looseness=-1

\subsubsection{Results}
The results suggest that our system can produce story outputs that are rated more favorably than those from the baseline in terms of both characterization and overall story quality.

\namedparagraph{Quantitative Results}
In the pairwise comparison by the model, stories generated by our system were preferred in 72 out of 100 cases.
The paired t-test on the model scores also showed a significant difference in favor of our system ($M = 6.24$ vs. $5.74, t = 6.52, p < .001^{***}$).
The human validation showed a similar pattern.
Using majority voting across the three raters, our system's stories were preferred for richer characterization in 96.7\% of pairs (29 out of 30) and for overall quality in 83.3\% of pairs (25 out of 30), with no pairs favoring the baseline under majority vote.
When pooling all individual judgments ($N = 90$), raters preferred our system's stories for characterization in 84.4\% of cases and for overall quality in 73.3\% of cases.\looseness=-1

\namedparagraph{Expert Comments}
In addition to evaluating the stories, experts shared the reasons for their judgments.
Raters consistently noted that stories from our approach provided ``\textit{specific, physical characterization}'' through concrete behavioral details---characters whose ``\textit{fingers often trembling as she fidgeted with her necklace,}'' whose ``\textit{knees itched where he'd knelt too long,}'' or who arrived ``\textit{still wearing his coffee shop apron.}''
Our stories also ``\textit{embodied}'' emotional changes rather than ``\textit{summarizing}'' them, allowing readers to feel ``\textit{how it felt}'' rather than simply being ``\textit{told what they did.}''
These details were seen as transforming characters from ``\textit{functional role-players}'' into ``\textit{vividly specific people,}'' giving them more ``\textit{texture}.''
Most of these details appeared to originate from the intermediate first-person perspectives in our pipeline.
However, in the minority of cases where raters preferred the baseline for overall quality, they cited its advantages in conciseness and narrative flow.
Raters described baseline stories as having ``\textit{cleaner sentence structure}'' and ``\textit{better pacing that holds the tension,}'' suggesting that the added characterization detail could occasionally come at the cost of readability, with our stories sometimes feeling ``\textit{bogged down by overwriting.}''\looseness=-1

%% file: sections/6-user-study.tex
\subsection{Study 2: User Evaluation}
Study~1 provided preliminary evidence that our pipeline can produce promising story outputs compared with a vanilla LLM baseline. However, output quality alone does not explain how writers experience the system or whether its interaction design supports characterization during writing. We therefore conducted a within-subjects study with 12 writers of different levels of creative writing expertise to investigate how \tool supports characterization during generative creative writing.
Specifically, we examined how \tool shapes writers' experiences of \textbf{controlling}, \textbf{manifesting}, and \textbf{tracing} characterization.


\subsubsection{Method}
Here, we describe the method of the study, including the baseline, participants, procedure, and analysis.

\namedparagraph{Baseline}
\revision{Recent HCI systems for characterization in story writing, such as CharacterChat~\cite{schmittCharacterChatSupportingCreation2021} and CharacterMeet~\cite{qinCharacterMeetSupportingCreative2024}, use chatbots to role-play story characters in support of character construction. 
Similarly, our baseline included character sheets and character chatbots for defining and interacting with characters (similar to CharacterChat and CharacterMeet).
In addition, to reflect common chatbot-based writing tools such as ChatGPT, our baseline also provided a story outline for plot conditioning and a story chatbot for generation and ideation. 
Example screenshots of the baseline are shown in \figref{fig:baseline}.
}
\looseness=-1

\namedparagraph{Participants}
We recruited 12 participants (8 female and 4 male), aged 25--65 (\(M = 38.92\), \(SD = 15.13\)), through crowdsourcing platforms, social networks, and word of mouth.
All participants reported proficiency in reading and writing in English.
We recruited participants with a range of creative writing expertise: 3 identified as professional writers with published work, 4 as advanced writers (2 of whom had also published work), 1 as an intermediate writer, and 4 as beginner writers.
We indicate participants' self-reported writing expertise when quoting them in the qualitative results.
Detailed information about each participant's prior writing experience, including relevant roles, genres, projects, and publications, is provided in Appendix~\ref{appendix:participant_information}.
In addition, all participants reported prior experience using AI tools for writing.
Their self-reported familiarity with using AI tools to support writing, measured on a 5-point scale (1 = none, 5 = extensive), was 3.92 (\(SD = 1.08\)).
We complemented these self-reports with participants' textual descriptions of their experience using AI tools for writing, including the tasks and purposes for which they had used them; these descriptions are also provided in Appendix~\ref{appendix:participant_information}.
We compensated each participant with \$20.\looseness=-1

\namedparagraph{Procedure}
The study began with informed consent\footnote{The study received approval from our institution's IRB.} and a demographics questionnaire.
Participants then completed two 30-minute writing sessions, each based on a different writing prompt (Appendix~\ref{appendix:writing_prompts_user}), one with our system and one with the baseline.
The order of systems and prompts was counterbalanced across participants.
Each session began with a 3--5-minute tutorial covering the key features of the assigned system.
Participants were then asked to create a three-act~\cite{fieldScreenplayFoundationsScreenwriting2005} outline based on the prompt, then using the assigned system to turn the outline into a story they are satisfied with.
Participants were encouraged to focus on characterization.
After each writing session, participants completed standardized post-condition measures, including the Creativity Support Index~\cite{cherryQuantifyingCreativitySupport2014} and the AI System Experience survey~\cite{wuAIChainsTransparent2022}, in 7-point Likert scale.
Following both conditions, participants completed a comparative questionnaire assessing which system better supported characterization. This included questions regarding controllability over character development, manifestation of character traits in the story, and traceability between characterization work and story text.
The study concluded with a 15-minute semi-structured interview to gather qualitative reflections on participants' experiences across the two conditions (questions are listed in Appendix~\ref{appendix:interview_questions}).
The entire study lasted approximately 90 minutes per participant.
\looseness=-1

\namedparagraph{Analysis}
For quantitative measures in the post-condition standardized surveys, we employed the Wilcoxon signed-rank test to account for the small sample size and the non-normal distribution of the data. 
To analyze the exit comparative questionnaire, we conducted a one-sample Wilcoxon signed-rank test using the neutral rating (4) as the population mean following prior work~\cite{processgalley_yen_2024}.
For the qualitative analysis of interview transcripts, we followed established thematic analysis protocols~\cite{braunUsingThematicAnalysis2006,scupinKJMethodTechnique1997} to identify emerging topics. 
The entire research team collectively reviewed the coding outcomes to refine the high-level themes.\looseness=-1

\input{figures/fig-interactionlog}

\subsubsection{Results}
We begin with an overview of user interaction patterns derived from logged events. 
This is followed by quantitative results from post-condition standardized surveys. 
Finally, we present quantitative results from the exit comparative survey regarding the control, traceability, and manifestation of characterizations, accompanied by qualitative user comments on each aspect.\looseness=-1

\namedparagraph{Interaction Patterns}
As illustrated in \figref{fig:interaction_log}, participants showed a consistent workflow across four primary activity phases when using our system: planning the outline, defining character arcs, generating first-person perspectives and selecting evidence, and finally generating and editing the third-person narrative. 
This aligns with our system's designed workflow. 
\revision{We then examined how participants used our system and the baseline differently as both systems supported staged writing from outline to narrative. 
We found that participants followed a similar workflow (from outlines to characters to narratives) in both conditions, but differed in how they controlled characterization: participants in baseline mainly revised global character personas, while participants using our system repeatedly edited keyframes at specific plot points.\looseness=-1

In addition, }time allocation when using our system across the phases in \figref{fig:interaction_log} varied depending on the user's writing expertise. 
Beginners dedicated the largest proportion of their time to the initial planning phase (42.94\%, compared to 29.90\% for Professionals and 23.04\% for Advanced users). Advanced writers, in contrast, invested heavily in designing character snapshots (49.26\%, compared to 27.55\% for Beginners and 28.03\% for Professionals). Professionals adopted a more balanced approach to early setup and spent the highest proportion of their time refining the final third-person narrative (31.29\%, compared to 17.82\% for Advanced users and 14.56\% for Beginners).
\looseness=-1

\input{tables/tab-survey}
\namedparagraph{Standardized Surveys}
\tabref{tab:survey} shows the quantitative results from the AI System Experience and CSI surveys. Overall, our approach provides a significantly more controllable, transparent, and engaging creative experience than the baseline.

\paragraph{AI System Experience}
Both systems were generally capable of meeting the task objectives ($M=6.58$ vs. $5.75$, $W=23.00, p=.072$).
However, participants rated \tool significantly higher than the baseline in terms of system transparency and cognitive support. Specifically, the system helped users better \textit{think through} the task ($M=6.92$ vs. $4.83, W=45.00, p=.004^{**}$) and was perceived as significantly more \textit{transparent} regarding its generative processes ($M=6.42$ vs. $4.00, W=55.00, p=.003^{**}$). Furthermore, users felt they had significantly more \textit{control} over the generation process when using \tool ($M=6.42$ vs. $5.33, W=37.00, p=.047^{*}$).
These benefits of transparency and controllability are further supported by our subsequent analysis of user ratings and comments regarding the affordances of our systems over controlling, manifesting, and tracing characterization.\looseness=-1

\paragraph{Creativity Support Index}
The results from the CSI indicate that \tool provided better support for generative creative writing workflows compared to the baseline. Users reported significantly higher levels of \textit{enjoyment} ($M=6.92$ vs. $5.75, W=28.00, p=.010^{**}$) and felt more \textit{immersed} in the activity ($M=6.25$ vs. $4.67, W=28.00, p=.011^{*}$). The system was also perceived as more \textit{worth the effort} required ($M=6.75$ vs. $6.08, W=15.00, p=.027^{*}$). Crucially for creative tasks, our system scored significantly higher in \textit{exploration} ($M=6.58$ vs. $4.58, W=45.00, p=.004^{**}$) and \textit{expressiveness} ($M=6.58$ vs. $5.08, W=45.00, p=.004^{**}$), suggesting that the our system allowed users to better explore the design space of characterization and express their creative intent in storytelling.\looseness=-1

\namedparagraph{Controlling Characterization}
We found that participants perceived \tool as offering stronger support for controlling characterization than the baseline. In the comparative survey, participants reported that it better supported their control over how each character's perspective was reflected in the final story ($M=6.00, SD=1.41, V=73.50, p=.003^{**}$), as well as their ability to deliberately shape each character's portrayal at different points in the story ($M=6.42, SD=0.79, V=78.00, p<.001^{***}$).



\paragraph{The Keyframed, Staged Workflow Enhances Controllability}
Participants said that \tool gave them a stronger sense of control by breaking characterization into a keyframed, staged workflow. Rather than relying on a single conversational thread, they could define character keyframes, inspect first-person narratives, and then select what should carry forward into the final story. This made the relationship between their inputs and the generated output feel more direct and predictable. P01 (Advanced) noted that ``\textit{your inputs had a direct bearing on the outcome}'' and appreciated being able to ``\textit{emphasize what you wanted... at each step of the process}'' and ``\textit{tweak it exactly how you want it to.}'' Participants contrasted this with the baseline, where control depended more on prompting and revising through chat. As P06 (Beginner) put it, ``\textit{Because everything is clearly separated and structured, ... the control of them is more direct.}''






\paragraph{Character Keyframes Make Character Arcs Explicit}
A major source of perceived control was the ability to define character states separately across plots. Participants said that the keyframing structure made character arcs explicit and editable, allowing them to shape not only who a character was, but how that character changed over plots. P07 (Professional) described using the system to create a stronger arc: ``\textit{I was able to see... how Monifa's character... she was, like, this passive character, but then in the third act, she got her backbone.}'' P02 (Professional) similarly appreciated being able to change a character internally, ``\textit{like, changing from selfish to selfless.}'' In contrast, participants noted that the baseline largely maintained a single persona unless they manually re-specified it. As P10 (Beginner) explained, ``\textit{you only have one version of the persona.}'' More broadly, participants valued being able to see and design ``\textit{the journey of the characters' traits through the acts}'' (P04, Professional). Some also found the interpolation feature useful for scaffolding transitions between character states.
For example, P10 (Beginner) said that when they knew where a character should begin and end, the interpolated middle state ``\textit{really speeds up the scaffolding.}''

\paragraph{Providing Fine-Grained Control Through Selection and Emphasis}
Participants also experienced control through the system's selection mechanisms. After generating first-person perspectives, they could choose which traits and evidence to emphasize in the final third-person story. This let them move beyond simply accepting or rejecting generated text, and instead curate what aspects of characterization should carry forward. P02 (Professional) emphasized the flexibility of keeping, discarding, and highlighting AI-generated character features: ``\textit{Because I can keep it but not use it, I can discard it so it's gone entirely, and then I can highlight one or two for each one.}'' For several participants, this emphasis mechanism functioned as a concrete lever of agency. As P08 (Beginner) explained, ``\textit{I can select which evidence or characteristics, I don't want to emphasize, or I want to emphasize. So... that give me more of control and agency.}'' P04 (Professional) summarized this difference succinctly: ``\textit{The huge difference is having the ability to choose which aspects of the character to focus on in the story.}'' In this sense, \tool supported control both by helping participants define characters, and by helping them decide what should matter most in the final narrative.\looseness=-1






\namedparagraph{Manifesting Characterization}
We found that participants preferred \tool over the baseline for manifesting characterization in the story. 
In the comparative survey, participants rated \tool significantly higher both for helping them incorporate character details that enriched characterization in the story (\(M=5.50, SD=1.93, V=66.00, p=.016^{*}\)) and for helping them translate abstract character ideas into concrete story details in the narrative (\(M=5.67, SD=1.16, V=55.00, p=.003^{**}\)).



\paragraph{Perspective Keyframes Help Concretize Characterization}
Participants valued first-person perspectives as an intermediate representation that made abstract character ideas more concrete before final story generation. By inspecting how the AI rendered a character's personality, motivations, and emotions through that character's own voice, they could assess whether the intended characterization had been realized and refine mismatches when needed. As P06 (Beginner) explained, the first-person perspective ``\textit{gives me a good window about how the AI understands my description of their personality. So if I see some mismatch, I can then refine my character properties. It's a good iteration process to both help me refine my thoughts and help me refine AI's thoughts.}'' Participants also appreciated that these perspectives manifested traits through concrete narrative forms such as description, dialogue, and emotion. P02 (Professional) said, ``\textit{I really like seeing how a character trait manifested itself in narrative, or dialogue, or expressed emotions of a character in any scene,}'' while P07 (Professional) noted that the system did ``\textit{a really good job}'' rendering characterization through sensory details such as ``\textit{her heels clacking.}'' Together, these first-person narratives made characterization feel less like a list of trait words and more like lived, narratively grounded behavior.

\paragraph{Perspective Keyframes Provide Reusable Materials for Story Generation}
Participants also described first-person perspectives as a rich pool of material that they could draw from when composing the final story. Instead of asking the AI to directly generate a third-person narrative from sparse character descriptions, they could first generate a fuller first-person account and then selectively carry forward the parts they wanted to emphasize. P05 (Beginner) explained, ``\textit{I think it's better to first write a complete and exhaustive first-person perspective, such that you can first ensure that what you are putting down is actually reflecting what you are planning with the character.}'' The same participant later described these outputs as ``\textit{kind of a cast of assets I can use in the final story,}'' from which they could choose passages to reflect what they truly wanted to emphasize. 
P01 (Advanced) similarly noted that these first-person perspectives were especially useful as a starting point for third-person omniscient writing, because they provided a base that participants could adapt and build on with the voices of different characters.\looseness=-1

\namedparagraph{Tracing Characterization}
We found that participants preferred \tool over the baseline for tracing characterization in the story. 
In the comparative survey, participants rated it significantly higher both for supporting their ability to trace how specific character traits were reflected in the final story (\(M=6.25, SD=1.06, V=66.00, p=.001^{**}\)) and for helping them identify which parts of the final story expressed the character attributes they intended (\(M=6.17, SD=1.40, V=64.50, p=.002^{**}\)).



\paragraph{Visible Mappings Help Writers Trace Traits into Text}
Participants valued \tool's explicit mappings between character traits, first-person perspectives, and final story text. Instead of inferring whether a trait had been reflected in the generated writing, they could directly inspect the connection through highlights and visual links. P02 (Professional) described this as a ``\textit{one-to-one correspondence,}'' explaining, ``\textit{Whatever I had selected from the AI auto output, or whatever I typed in, it connected those two directly.}'' This visibility also reduced the burden of manually searching for evidence in the text. As P06 (Beginner) noted, the system made it easier to verify that intended traits were actually reflected in the generated paragraphs, rather than having to find that evidence independently. Participants also appreciated visual encodings such as color, which made traceability easier to perceive at a glance. P12 (Advanced) remarked, ``\textit{I think humans respond well to color... it helps with the transparency.}'' Together, these features made characterization traceable both conceptually and perceptually through the interface itself.\looseness=-1

\paragraph{Traceability Supports Reflection on Writing Decisions}
Traceability also helped participants reflect on how their characterization decisions shaped the generated story.
Participants described gaining a clearer understanding of how the system transformed their inputs into narrative language, which in turn made the writing process feel more meaningful and intentional.
P02 (Professional) explained, ``\textit{I'm able to understand, oh, there's a one-to-one relationship between the specific things I put for the character... And then that is specifically output in the story, and the software allows me to visually see those two together, and that allows my brain to make the connection between my chosen description and the language model's manipulation of that into a narrative that matches all the stuff that I typed.}'' 
This visibility appeared to support a more reflective and deliberate mode of authorship. 
P02 contrasted the experience with less structured forms of writing, noting, ``\textit{I think writers who just start writing and just do stream of consciousness don't ever understand why they're putting anything down. Whereas this allowed me to see that.}'' 
The same participant later connected this traceability to a stronger sense of human contribution: ``\textit{I did so much high-level thinking, making decisions, which is what us humans are supposed to do.}'' This visibility supported a more reflective and deliberate mode of authorship, helping participants see themselves as actively shaping characterization rather than merely reacting to generated text.\looseness=-1

%% file: figures/fig-interactionlog.tex
\begin{figure}
  \centering
  \includegraphics[width=\linewidth]{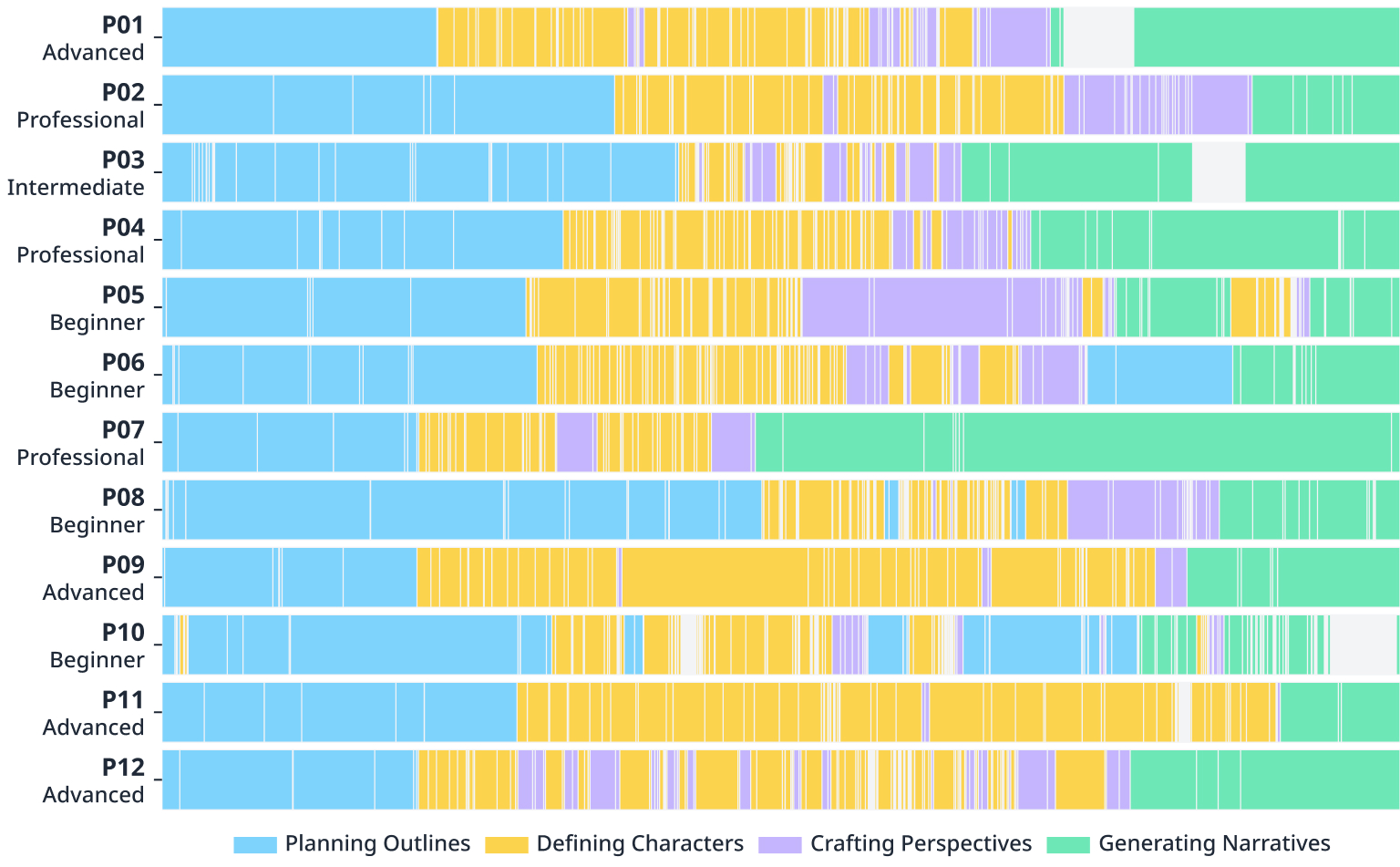}
  \caption{\formatcaption{\revision{Overview of participants' time distribution.}}{Each row represents a single participant, labeled by their creative writing expertise. The x-axis tracks normalized time as a percentage of task progress. Colored segments denote specific activity categories: \emph{Planning Outlines} (drafting the story outline), \emph{Defining Characters} (creating character keyframes), \emph{Crafting Perspectives} (generating perspective keyframes and selecting evidence), and \emph{Generating Narratives} (generating, reviewing, and revising narratives). The width of each segment reflects the relative time spent on that activity. \revision{Participants completed the three-act story writing task with our system in 25.92 mins on average.}}}
  \Description{A horizontal stacked timeline chart summarizes how 12 participants distributed their time across the writing task. Each row corresponds to one participant, labeled P01 through P12, with a second label indicating expertise level: Advanced, Professional, Intermediate, or Beginner. Time progresses from left to right as normalized task progress, so each row spans the full duration of that participant’s session. The rows are divided into many narrow colored segments representing four activity categories: Planning Outlines, Defining Characters, Crafting Perspectives, and Generating Narratives. Lighter blue segments correspond to earlier planning and character work, while darker blue segments indicate perspective writing and final narrative generation. Most participants begin with a long stretch of Planning Outlines, then transition into more mixed alternation between Defining Characters and Crafting Perspectives, and many end with longer blocks of Generating Narratives. However, the patterns vary across participants: some, such as P07, shift earlier into extended narrative generation, while others, such as P10, spend a large portion of the later session alternating among perspective and narrative-related activities. Overall, the figure shows that participants followed a shared staged workflow but differed in how much time they devoted to each phase and in how often they switched between activities.}
  \label{fig:interaction_log}
\end{figure}

%% file: tables/tab-survey.tex
\begin{table}
\footnotesize
\centering
\caption{\formatcaption{Survey results of perceived experience on AI systems~\cite{wuAIChainsTransparent2022} and Creativity Support Index (CSI)~\cite{cherryQuantifyingCreativitySupport2014} under two conditions.}{Wilcoxon signed-rank paired t-test W-values and p-values (*: \(p<.05\), **: \(p<.01\), ***: \(p<.001\)) are reported. Like previous work~\cite{massonVisualStoryWritingWriting2025,suhLuminateStructuredGeneration2024}, we omitted the Collaboration factor to avoid confusion, as the tasks did not involve human collaboration.}}
\label{tab:survey}
\begin{tabular}{lcccccl}
\toprule
\multirow{2}{*}{\textbf{Scales}}
& \multicolumn{2}{c}{\textbf{Ours}}
& \multicolumn{2}{c}{\textbf{Baseline}}
& \multicolumn{2}{c}{\textbf{Statistics}} \\
\cmidrule(lr){2-3}
\cmidrule(lr){4-5}
\cmidrule(lr){6-7}
& \textbf{M}
& \textbf{SD}
& \textbf{M}
& \textbf{SD}
& \textbf{W}
& \textbf{\textit{p}} \\
\midrule

\multicolumn{7}{l}{\textbf{AI System Experience}} \\
\quad Match Goal
  & 6.58 & 1.17 & 5.75 & 1.66 & 23.00 & .072 \\
\quad Think Through
  & 6.92 & 0.29 & 4.83 & 2.08 & 45.00 & .004** \\
\quad Transparent
  & 6.42 & 0.79 & 4.00 & 2.17 & 55.00 & .003** \\
\quad Controllable
  & 6.42 & 1.44 & 5.33 & 1.44 & 37.00 & .047* \\

\midrule
\multicolumn{7}{l}{\textbf{Creativity Support Index}} \\
\quad Enjoyment
  & 6.92 & 0.29 & 5.75 & 1.36 & 28.00 & .010** \\
\quad Immersion
  & 6.25 & 0.97 & 4.67 & 1.78 & 28.00 & .011* \\
\quad Worth Effort
  & 6.75 & 0.45 & 6.08 & 1.31 & 15.00 & .027* \\
\quad Exploration
  & 6.58 & 0.52 & 4.58 & 2.02 & 45.00 & .004** \\
\quad Expressiveness
  & 6.58 & 0.90 & 5.08 & 1.83 & 45.00 & .004** \\
\bottomrule
\end{tabular}
\Description{A seven-column table compares survey results for the authors’ system and a baseline across measures of AI system experience and creativity support. For each scale, the table reports the mean and standard deviation for both systems, along with a Wilcoxon signed-rank test statistic and p-value. In the AI System Experience section, the authors’ system scores higher than the baseline on all four measures: Match Goal, Think Through, Transparent, and Controllable. The difference is statistically significant for Think Through, Transparent, and Controllable, but not for Match Goal. The largest differences appear in Think Through, where the authors’ system has a mean of 6.92 versus 4.83 for the baseline, and Transparent, where it has a mean of 6.42 versus 4.00. In the Creativity Support Index section, the authors’ system also scores higher on all five measures: Enjoyment, Immersion, Worth Effort, Exploration, and Expressiveness. All five differences are statistically significant. The largest differences appear in Exploration, with means of 6.58 versus 4.58, and in Expressiveness, with means of 6.58 versus 5.08. Overall, the table shows that participants rated the authors’ system more positively than the baseline on nearly every reported dimension, with significant advantages in most measures related to transparency, reflection, exploration, and expressive support.}
\end{table}

%% file: sections/7-discussion.tex
\section{Discussion}

\subsection{Design Implications}

\subsubsection{Characterization as an Evolving Creative Process}
Prior work often treats characterization as something specified upfront, for example through character descriptions, personas, or profiles that are then used to guide later generation stages~\cite{parkCharacterCentricCreativeStory2025,schmittCharacterChatSupportingCreation2021,qinCharacterMeetSupportingCreative2024,mirowskiCoWritingScreenplaysTheatre2023}. 
Our findings suggest that characterization is better understood as an evolving creative process that develops in at least two directions: horizontally across story plots, as characters change over events, and vertically across perspectives, as those changes are explored through different points of view.
Rather than collapsing characterization into a single prompt or profile, AI writing systems can better support this process by exposing intermediate representations that help writers plan, inspect, and refine character development before generating final prose.


\subsubsection{Perspective as a Design Material for Writing Tools}
Prior work has used role-play to let AI enact user-defined characters and help writers refine character profiles before writing~\cite{fuYourStoryCoCreative2025,schmittCharacterChatSupportingCreation2021,qinCharacterMeetSupportingCreative2024,sunORIBAExploringLLMDriven2025,wangStoryVerseCoauthoringDynamic2024}. 
Our findings suggest a broader role for perspective in AI writing tools.
In our system, first-person perspectives served as an intermediate representation that helped writers concretize abstract character ideas, inspect how traits might be expressed in language, and reason across multiple characters' inner lives. This suggests that perspective is not only useful for conversational exploration, but can also function as a reusable design material that bridges character planning and story generation.


\subsubsection{Traceability Matters for Generative Creative Writing}
Our findings suggest that traceability is an important property of AI writing tools. By making visible how selected traits and evidence were reflected in generated text, our system helped writers inspect whether their intentions were carried into the final story and understand how character-related inputs were transformed across stages of the workflow. This was valuable for both verification and reflection: participants used these mappings to check what had been emphasized, compare generated text with their intentions, and refine characterization decisions accordingly. This aligns with prior work showing the value of provenance for transparent AI-assisted writing~\cite{hoqueHaLLMarkEffectSupporting2024,siddiquiDraftMarksEnhancingTransparency2025}. More broadly, this suggests that AI writing tools should make connections between writer-authored materials and generated outputs explicit, especially when supporting creative work that unfolds through multiple stages.

\subsection{Limitations and Future Work}
Our system has several limitations.
First, our system currently supports a relatively structured workflow. It may feel restrictive for writers who prefer to develop stories in a more fluid or improvisational way. Future work could explore more flexible interactions that allow writers to move between planning, character exploration, and drafting in a less linear manner.
\revision{
Additionally, although keyframing and first-person perspective keyframes can extend to longer stories, our current design does not yet cleanly support highly non-linear plots, such as second-act flashbacks, which are interesting directions for future work.
Lastly, our current instantiation of \tool focuses primarily on characterization through character and perspective keyframes. Future work could explore keyframing other evolving narrative properties, such as tone, pacing, or scenes, as well as interactions for coordinating multiple property tracks within the same story.}
\looseness=-1

Our study also has several limitations that future work could address.
First, the study included 12 participants, which is a relatively small sample size.
Although we conducted statistical tests, we do not treat these results as conclusive; instead, they should be interpreted as promising but preliminary.
\revision{Second, due to time constraints, the writing task asked participants to produce a short story using a three-act structure.
Future work should examine how our system supports longer-form story writing.}
Third, our evaluation focused on a controlled writing task with two prompts and a comparison against one chatbot-based baseline.
Future work could test the system with a broader range of writing goals and baseline conditions.\looseness=-1

%% file: sections/8-conclusion.tex
\section{Conclusion}
In this paper, we introduce \tool, a new interaction design for generative creative writing that uses high-level representations of plot events, character arcs, and narrative perspective to help writers plan and guide story generation. We instantiate this design in an interactive system that connects high-level story planning to final narrative generation through plot, character, and perspective keyframes, supporting writers in shaping character arcs, exploring first-person expression, and guiding third-person storytelling. Through a technical evaluation and a user study, we find that our approach produces stories with higher overall quality and richer characterization, while also supporting a more controllable, transparent, and engaging writing experience than a baseline system that reflects the current design of generative creative writing tools.
More broadly, this work shows how keyframing can serve as an interaction paradigm for human-AI co-writing by helping writers balance automation with creative control.\looseness=-1

%% file: sections/z-appendix.tex
\section{Implementation Details}
\label{appendix:technical_details}

In this section, we present the prompts used to instruct GPTs to
suggest character traits, interpolate character keyframes, generate
first-person perspectives, extract evidence from perspectives, and
generate narratives based on selected traits and evidences.

\subsection{Suggesting Physiology Traits}

\begin{myfancybox}
\footnotesize

\noindent\textbf{Physiology:}
{\ttfamily\raggedright\obeylines
You are a story development assistant.
\bigskip
Full story: \promptbox{<full\_outline>}
Targeting plot: \promptbox{<current\_plot>}
Character: \promptbox{<character\_name>}
Existing physiology traits: \promptbox{<existing\_traits>}
\bigskip
Brainstorm three concise physiology traits for this targeting plot in the story. These traits should sharpen characterization and help guide revisions to improve the story.
\bigskip
Guidelines:
- Focus on Physical appearance, clothing, body language, visible characteristics.
- Keep each trait 3-8 words
- Ground traits in the story plot and the full story
- Avoid repeating existing traits or near-duplicates
- Return exactly three traits
\bigskip
Return JSON that matches the provided schema.
}
\end{myfancybox}

\subsection{Suggesting Psychology Traits}

\begin{myfancybox}
\footnotesize

\noindent\textbf{Psychology:}
{\ttfamily\raggedright\obeylines
You are a story development assistant.
\bigskip
Full story: \promptbox{<full\_outline>}
Targeting plot: \promptbox{<current\_plot>}
Character: \promptbox{<character\_name>}
Existing psychology traits: \promptbox{<existing\_traits>}
\bigskip
Brainstorm three concise psychology traits for this targeting plot in the story. These traits should sharpen characterization and help guide revisions to improve the story.
\bigskip
Guidelines:
- Focus on Emotions, motivations, thoughts, beliefs, mental state.
- Keep each trait 3-8 words
- Ground traits in the story plot and the full story
- Avoid repeating existing traits or near-duplicates
- Return exactly three traits
\bigskip
Return JSON that matches the provided schema.
}
\end{myfancybox}

\subsection{Suggesting Sociology Traits}

\begin{myfancybox}
\footnotesize

\noindent\textbf{Sociology:}
{\ttfamily\raggedright\obeylines
You are a story development assistant.
\bigskip
Full story: \promptbox{<full\_outline>}
Targeting plot: \promptbox{<current\_plot>}
Character: \promptbox{<character\_name>}
Existing sociology traits: \promptbox{<existing\_traits>}
\bigskip
Brainstorm three concise sociology traits for this targeting plot in the story. These traits should sharpen characterization and help guide revisions to improve the story.
\bigskip
Guidelines:
- Focus on Social roles, relationships, status, interactions with others.
- Keep each trait 3-8 words
- Ground traits in the story plot and the full story
- Avoid repeating existing traits or near-duplicates
- Return exactly three traits
\bigskip
Return JSON that matches the provided schema.
}
\end{myfancybox}

\subsection{Interpolating Character Keyframes}

\begin{myfancybox}
\footnotesize

{\ttfamily\raggedright\obeylines
You are analyzing a character's narration to extract their traits at this specific plot in the story.
\bigskip
Narration of this plot by \promptbox{<charactor\_name>}:
\promptbox{<perspective\_of\_this\_plot>}
\bigskip
\textnormal{\textit{If nearby snapshots exist:}}
Character keyframes from nearby plots:
\promptbox{<Previous|Following>} keyframe of \promptbox{<keyframe\_name>}:
\quad Physiology: \promptbox{<trait1, trait2, ...>}
\quad Psychology: \promptbox{<trait1, trait2, ...>}
\quad Sociology: \promptbox{<trait1, trait2, ...>}
\bigskip
\textnormal{\textit{If first-person perspective is available:}}
Full narration (context only --- do NOT quote from this section):
\promptbox{<full\_perspective\_text>}
Based on the narration text\promptbox{< and nearby keyframes>}, extract the character traits for \promptbox{<character\_name>} at this plot.
\bigskip
Guidelines:
- Physiology: Physical appearance, clothing, body language, visible characteristics
- Psychology: Emotions, motivations, thoughts, beliefs, mental state
- Sociology: Social roles, relationships, status, interactions with others
- Consider the character's development trajectory from nearby keyframes (if any)
- Only include traits that are evident or strongly implied in the text
- Keep trait descriptions concise (3-8 words each)
- Return 2-5 traits per category when evident
\bigskip
Evidence requirements:
- For every trait you include, add one entry to traitEvidence with the traitCategory, the exact trait wording, and an evidenceText.
- Each evidenceText must be a verbatim quote from the narration above (do NOT pull from the context section).
- If you cannot find a direct quote, omit the trait entirely.
\bigskip
Return JSON that matches the provided schema.
}
\end{myfancybox}

\subsection{Generating First-Person Perspectives}

\begin{myfancybox}
\footnotesize

{\ttfamily\raggedright\obeylines
You are a story writer. Your job is to write a first-person narration for a given story plot from the perspective of a specified character.
\bigskip
Requirements:
- Stay faithful to the facts, chronology, and causality in the Targeting plot and the Full story.
- Do not contradict established details (names, places, outcomes, revealed secrets, injuries, timelines, motivations already shown).
- Do not add new major plot events; you may add small, plausible sensory details and moment-to-moment actions that do not change the plot's outcome.
- Keep tense and POV consistent: first-person ("I", "me", "my").
- Show emotions and thinking through actions, speech, appearance, environment, and specific observations (avoid generic statements like "I was scared" unless grounded in concrete detail).
- Match the tone and genre implied by the Full story.
- Strict length limit: Maximum 200 words.
\bigskip
Character voice rules:
- If character traits are supplied, demonstrate those traits through concrete choices in diction, focus, and interpretation (what they notice, what they ignore, how they justify things).
- If no traits are provided, infer a character-consistent voice from the Full story and Targeting plot.
\bigskip
Return each result as a JSON object that satisfies the provided schema.
\bigskip
Full story:
\promptbox{<plot[0]>}
\promptbox{<plot[1]>}
\promptbox{<\ldots>}
\bigskip
Targeting plot:
\promptbox{<targeting\_plot>}
\bigskip
Narrator: \promptbox{<charactor\_name>}
\textnormal{\textit{If a character keyframe with traits exist for this plot:}}
Character traits:
- Physiology: \promptbox{<trait1, trait2, ...>}
- Psychology: \promptbox{<trait1, trait2, ...>}
- Sociology: \promptbox{<trait1, trait2, ...>}
\bigskip
\textnormal{\textit{If no character keyframe for this plot:}}
Character traits: (none provided)
\bigskip
\textnormal{\textit{If custom prompt is provided:}}
ADDITIONAL INSTRUCTIONS FROM USER:
\promptbox{<custom\_prompt>}
}
\end{myfancybox}

\subsection{Extracting Evidence from Perspectives}

\begin{myfancybox}
\footnotesize

{\ttfamily\raggedright\obeylines
You are an expert literary analyst. Identify direct textual evidence (i.e., verbatim phrases) that confirms the given character traits.
\bigskip
Full story (background only---do NOT quote from this section):
\promptbox{<group\_context>}
Current plot (ONLY source for evidence):
\promptbox{<reflection>}
Characters and traits to verify:
\promptbox{<character\_name>}
\quad Physiology:
\quad\quad - \promptbox{<trait\_value>}
\quad\quad - \promptbox{<trait\_value>}
\quad Psychology:
\quad\quad - \promptbox{<trait\_value>}
\quad Sociology:
\quad\quad - \promptbox{<trait\_value>}
\bigskip
Evidence categories to classify each phrase:
- directDefinition: Explicit direct statements or labels about the character
- actions: Physical actions, behaviors, or body language
- speech: What the character says, how they speak, or how other characters say about them
- appearance: Visual descriptions of the character
- environment: Surroundings, context, or setting that characterizes the person
\bigskip
Instructions:
1. Scan the current snippet for exact short phrases that directly or indirectly demonstrate each listed trait.
2. Only report evidence that appears verbatim in the current snippet text.
3. When one phrase supports multiple traits from the same category, list all matching traits together.
4. Assign each phrase to exactly one evidence category from the list above.
5. Return characterEvidence entries in the same order as the character list above.
6. Return JSON that matches the provided schema exactly. Do not include explanations outside the schema.
}
\end{myfancybox}

\subsection{Generating Third-Person Narratives}

\begin{myfancybox}
\footnotesize

\noindent\textbf{Generating Third-Person Narratives:}
{\ttfamily\raggedright\obeylines
You are a narrative writer. Expand the provided story outline into a third-person story.
\bigskip
Story outlines:
\promptbox{<plot[0]>}
\promptbox{<plot[1]>}
\promptbox{<\ldots>}
Main characters:
\promptbox{<character\_list>}
\bigskip
Instructions:
- Write a cohesive full story that follows the outline exactly
- Use third-person narration
- Include both main characters throughout
- Maintain chronological order and clear act progression
- Return the story per act, in order, with 1-2 paragraphs per act
- Each act entry should include the act number, the act label from the outline, and the act text
\bigskip
Return JSON that matches the provided schema.
}
\end{myfancybox}

\begin{myfancybox}
\footnotesize

\noindent\textbf{Enriching with Selected Traits and Evidence:}
{\ttfamily\raggedright\obeylines
You are a narrative editor. Your job is to make the original story read *better* by seamlessly integrating the selected details.
\bigskip
Original story + selected details:
Plot 1: \promptbox{<plot\_description>}
- Character: \promptbox{<character\_name>}
- Traits: \promptbox{<trait1, trait2, ...>}
\bigskip
\textnormal{\textit{If snippets exist for this plot:}}
Selected details:
\quad 1. "\promptbox{<evidence\_from\_perspectives>}"
\quad 2. "\promptbox{<evidence\_from\_perspectives>}"
\bigskip
\textnormal{\textit{If no snippets:}}
Selected details: (none)
Plot 2: \promptbox{<plot\_description>}
\promptbox{<\ldots>}
\bigskip
Requirements:
- Preserve the original plot, beat order, and third-person narration.
- Do NOT add new events, attempts, or outcomes beyond what the original story already includes.
- Integrate details naturally (avoid "laundry lists" of descriptions).
- Avoid overwriting: keep sentences clear and varied in length; do not let any one sentence run on too long.
- Maintain continuity (names, timelines, locations, and cause-and-effect must remain consistent).
\bigskip
Snippet usage tracking:
For each plot with selected character details, output "snippetUsages" as pairs of:
\quad - originalSnippet: exact text from selected details (first-person)
\quad - verbatimInNarrative: an EXACT substring from your third-person narrative showing your transformation (\ensuremath{\leq}25 words unless impossible)
\quad For events without details, snippetUsages must be an empty array.
\bigskip
Output:
- Return JSON matching the provided schema.
\bigskip
\textnormal{\textit{If custom prompt is provided:}}
ADDITIONAL INSTRUCTIONS FROM USER:
\promptbox{<custom\_prompt>}
}
\end{myfancybox}

\section{Evaluation Details}

\subsection{Writing Prompts in Technical Evaluation}
\label{appendix:writing_prompts_tech}
To generate stories, we used the ten writing prompts from the \texttt{CoAuthor} dataset~\cite{leeCoAuthorDesigningHumanAI2022a}, supplemented by another ten prompts randomly sampled from the \texttt{WritingPrompts} dataset~\cite{fanHierarchicalNeuralStory2018}.
The full list of the 20 writing prompts are shown below:

\begin{enumerate}
    \item Once upon a time there was an old mother pig who had one hundred little pigs and not enough food to feed them. So when they were old enough, she sent them out into the world to seek their fortunes. You know the story about the first three little pigs. This is a story about the 92nd little pig. The 92nd little pig built a house out of depleted uranium. And the wolf was like, ``dude.''

    \item A woman has been dating guy after guy, but it never seems to work out. She's unaware that she's actually been dating the same guy over and over; a shapeshifter who's fallen for her, and is certain he's going to get it right this time.

    \item When you die, you appear in a cinema with a number of other people who look like you. You find out that they are your previous reincarnations, and soon you all begin watching your next life on the big screen.

    \item Humans once wielded formidable magical power. But with over 7 billion of us on the planet now, mana has spread far too thinly to have any effect. When hostile aliens reduce humanity to a mere fraction, the survivors discover an old power has begun to reawaken once again.

    \item An alien has kidnapped Matt Damon, not knowing what lengths humanity goes through to retrieve him whenever he goes missing.

    \item You're Barack Obama. Four years into your retirement, you awake to find a letter with no return address on your bedside table. It reads, ``I hope you've had a chance to relax, Barack\ldots but pack your bags and call the number below. It's time to start the real job.'' Signed simply, ``JFK.''

    \item Following World War III, all the nations of the world agreed to 50 years of strict isolation from one another in order to prevent additional conflicts. Fifty years later, the United States comes out of exile, only to learn that no one else went into isolation.

    \item Your entire life, you've been told you're deathly allergic to bees. You've always had people protecting you from them, be it your mother or a hired hand. Today, one slips through and lands on your shoulder. You hear a tiny voice say, ``Your Majesty, what are your orders?''

    \item All of the ``\#1 Dad'' mugs in the world change to show the actual ranking of dads suddenly.

    \item When you're 28, science discovers a drug that stops all effects of aging, creating immortality. Your government decides to give the drug to all citizens under 26, but you and the rest of the ``Lost Generations'' are deemed too high-risk. When you're 85, the side effects are finally discovered.

    \item A boy pretends he is an astronaut in order to help cope with concepts and situations he can't understand.

    \item By the time humans come along, elves had invented space travel, and dwarves had split the atom. One hundred years later, the world looks like your typical fantasy setting. How did it happen?

    \item A time traveller interviews major historical figures at three points in their lives: their 16th birthday, the day after they made their most important decision, and the day before they die.

    \item Everyone has superpowers, but the richer you are, the weaker your powers become.

    \item Every fifty years, the accumulated wealth of the world is randomly redistributed. Tonight is the eve of the global redistribution.

    \item A woman comes into the same diner every morning, orders the same meal, and always leaves without eating a bite.

    \item Due to a crossed line, a customer support worker has to deal with a hostage situation. Meanwhile a hostage negotiator has to deal with a disgruntled customer.

    \item Construction workers are exposed to a relic of magical power while beginning work on a new building. Slowly, it begins to change them\ldots

    \item A retired supervillain is in the bank with his 6-year-old daughter when a new crew of supervillains comes in to rob the place.

    \item Twin brothers with a strong telepathic connection discover the elixir of life. Only one is granted immortality, but their telepathic connection transcends the mortal brother's death, providing the first physical world/afterlife connection.
\end{enumerate}

\subsection{Writing Prompts in User Evaluation}
\label{appendix:writing_prompts_user}
Below are the two writing prompts used in the user evaluation.

\begin{enumerate}
    \item Two characters with very different personalities are forced to work together toward a difficult goal. At first they clash, but over time they must decide whether to trust each other. Write a story about how their relationship develops.
    \item Two characters who trust each other uncover a secret that could change their community. One wants to reveal it; the other wants to keep it hidden. Write a story about how their relationship and choices change as they face the consequences.
\end{enumerate}

\subsection{Baseline System Interface}
\label{appendix:baseline_interface}
\revision{
Our baseline system includes character sheets and character chatbots for defining and interacting with characters (similar to CharacterChat~\cite{schmittCharacterChatSupportingCreation2021} and CharacterMeet~\cite{qinCharacterMeetSupportingCreative2024}), as well as a story outline for plot conditioning and a story chatbot for generation and ideation. 
\figref{fig:baseline} shows an example screenshot of the baseline interface.
}

\input{figures/fig-baseline}

\subsection{Participant Information}
\label{appendix:participant_information}
through crowdsourcing platforms, social networks, and word of mouth. All participants reported proficiency in reading and writing in English. Participants had a range of creative writing expertise: 3 identified as professional writers with published work, 4 as advanced writers (2 of whom had also published work), 1 as an intermediate writer, and 4 as beginner writers. In addition, all participants reported prior experience using AI tools for writing. Their self-reported familiarity with using AI tools to support writing, measured on a 5-point scale (1 = none, 5 = extensive), was 3.92 (\(SD = 1.08\)).
Detailed participant information is shown in \tabref{tab:demographics}.\looseness=-1

\input{tables/tab-participants.tex}

\subsection{Interview Questions}
\label{appendix:interview_questions}

\begin{enumerate}
\item Looking across the two systems, how did your experience of developing and writing characters differ?
\item In which system did you feel more in control of how characterization was portrayed in the final story?
\item How did using 1st-person character perspectives as an intermediate step affect your understanding and expression of the character in the final story?
\item How did you decide when to use each of the three interfaces (canvas, track, and table)? For what kinds of tasks did you prefer each one, and why?
\item How, if at all, did either system affect your ability to deliberately plan character development throughout the story?
\item How did each system help or hinder you in translating abstract character ideas into concrete story details?
\item How easy was it, in each system, to trace aspects of the final story back to your intended character traits or earlier characterization work?
\item For characterization specifically, how would you use the system in your future writing practice?
\item What improvements would you make to the system to better support characterization in your future writing practice?
\end{enumerate}

%% file: figures/fig-baseline.tex
\begin{figure*}
  \centering
  \includegraphics[width=0.95\linewidth]{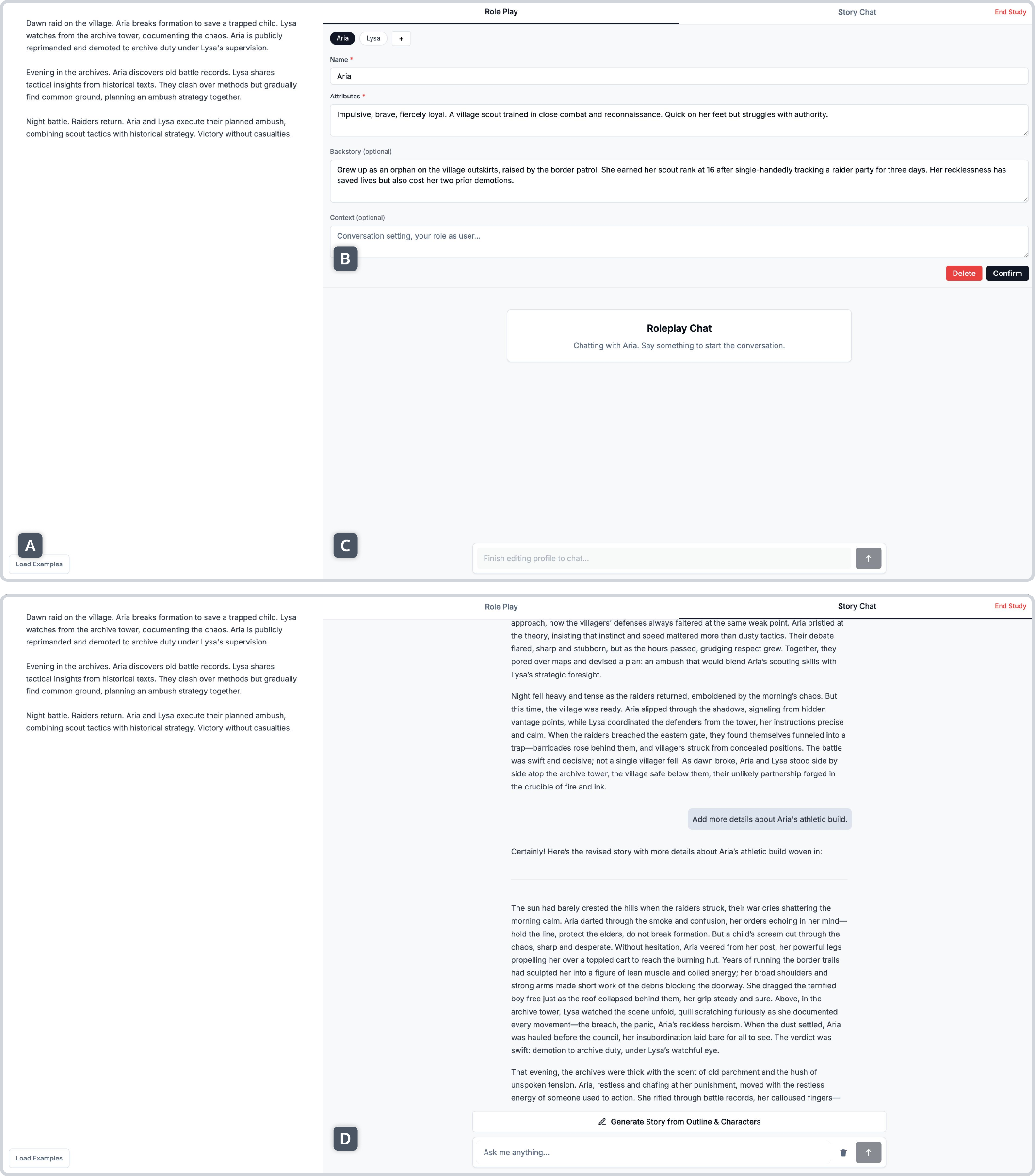}
  \caption{\revision{\formatcaption{Screenshots of the baseline system.}{The baseline system includes a story outline (A) for plot conditioning and a story chatbot (D) for generation and ideation, as well as character sheets (B) and character chatbots (C) for defining and interacting with characters (similar to CharacterChat~\cite{schmittCharacterChatSupportingCreation2021} and CharacterMeet~\cite{qinCharacterMeetSupportingCreative2024}).}}}
  \Description{Two screenshots show the baseline system for creative writing. In both screenshots, the interface is split into a narrow left panel containing a short three-event story outline and a much larger right panel for interaction. The top screenshot shows the character setup stage. On the right, a tabbed “Role Play” interface is open with character tabs for Aria and Lysa. Form fields let the writer enter a character name, attributes, backstory, and optional context. Below these fields is an empty roleplay chat area with a message indicating that the user can begin chatting with the selected character once profile editing is complete. The bottom screenshot shows the story generation stage. The same outline remains visible in the left panel, while the right panel now displays a story chat containing a long generated narrative passage. A user message asks to add more details about Aria’s athletic build, and the assistant responds with a revised version of the story incorporating that request. At the bottom of the chat is a button labeled “Generate Story from Outline & Characters” above a standard chat input box. Together, the two screenshots illustrate the baseline workflow: writers first define characters through form-based roleplay profiles, then iteratively prompt and revise the generated story through a single chat interface.}
  \label{fig:baseline}
\end{figure*}

%% file: tables/tab-participants.tex
\begin{table*}
\footnotesize
\centering
\caption{\formatcaption{Demographic information for participants.}{This table presents participants' ages, genders, creative writing experience, and use of AI tools for writing. We slightly modified their descriptions of writing experience and AI use to avoid identifiable information.}}
\label{tab:demographics}
\begin{tabular}{lllp{0.4\linewidth}p{0.4\linewidth}}
\toprule
\textbf{ID} & \textbf{Gender} & \textbf{Age} & \textbf{Creative Writing Experience} & \textbf{AI Usage Experience} \\
\midrule
P01 & Female & 49 & \textbf{\textit{Advanced:}} I published a speculative fiction young adult book recently. & \textbf{\textit{High:}} I've used ChatGPT for fact checking and to check spelling, grammar and comprehensibility. I've used Claude Sonnet 4.6 to check my stories for developmental weaknesses (character arcs, plot). \\
P02 & Male & 60 & \textbf{\textit{Professional:}} I am working on book 30 in a series of science fiction, printing all 29 previous books at The Book Patch. I have created hundreds of educational workbooks for teachers, many full of poems, stories, descriptions, word problems, etc. & \textbf{\textit{Extensive:}} I rely on Claude and ChatGPT to assist me before, during, and after writing. I set up the entire world within a book, using these two (and Gemini occasionally) to give me detailed explanations of the specific content each chapter will explore, how things are done (piloting a ship, digging out a gem, communicating with an alien species, dangers in space travel, etc.) and ways to allow my two main characters to experience awe, curiosity, wonder, focus, patience, etc.\\
P03 & Female & 28 & \textbf{\textit{Intermediate:}} I practiced writing stories for exams. & \textbf{\textit{High:}} I use ChatGPT for academic writing, such as polishing my content and helping with my thoughts.\\
P04 & Female & 33 & \textbf{\textit{Professional:}} I am currently working on a screenplay set during the American Revolution, The film follows the first black poet to be published in the US. & \textbf{\textit{Extensive:}} I use it a lot to check historical accuracies. I also use it for feedback on outlines because it gets me brainstorming.\\
P05 & Male & 25 & \textbf{\textit{Beginner:}} I enjoy the idea of writing stories, though I haven't explored it very much yet. & \textbf{\textit{Limited:}} I use AI tools to find the best word/phrase to describe something. I also use AI to re-write sentences when I feel a sentence sounds weird or tedious.\\
P06 & Female & 28 & \textbf{\textit{Beginner:}} I'm interested in writing fictions, but I haven’t tried it much yet. & \textbf{\textit{Limited:}} I've used ChatGPT, Claude, Gemini. I usually ask AI to help me rewrite emails.\\
P07 & Female & 55 & \textbf{\textit{Professional:}} I have a master's degree in creative writing and have published several horror and sci-fi short stories in literary journals. I sold a TV movie and one of my short stories was published in a New York Times best-selling anthology. & \textbf{\textit{High:}} I mainly use ChatGPT for help with my freelance clients, such as writing newsletters or bios. When I was interviewing for a position that involved writing verticals (micro stories), I asked ChatGPT to show me an example of a sci-fi vertical. I used Copilot to help me generate beats for a new screenplay.\\
P08 & Female & 25 & \textbf{\textit{Beginner:}} As a hobby, I wrote some short pieces, mostly short scenes. I used to write short science fictions when I was younger. & \textbf{\textit{Moderate:}} I used AI tools for academic writing a lot, especially paraphrasing and proofreading.\\
P09 & Female & 65 & \textbf{\textit{Advanced:}} I have written two ebooks and a screenplay (131 pages long) available for sale on Amazon Kindle. Most currently, I frequently write scripts and record them for audio entertainment. I have ghost written a few books, wrote a one hour script for a podcast, and won a national writing contest as a college student for film criticism. & \textbf{\textit{High:}} I just recently experimented with AI for writing. As a new Grandma, I asked for inspiration for a baby's storybook audio. To my surprise, about twenty minutes later, my AI had created a flawless nature themed personalized story.\\
P10 & Male & 29 & \textbf{\textit{Beginner:}} I've seldom been writing fiction and stories recently, but I do write user journeys for my products. & \textbf{\textit{High:}} ChatGPT and Claude are mostly used to help with scientific writing. They have been mostly very helpful.\\
P11 & Male & 46 & \textbf{\textit{Advanced:}} I've been writing raps for over 25 years as well as poetry. I also have written three fictional books but haven't published anything and I'm currently writing two more. & \textbf{\textit{Extensive:}} I have used Gemini, Chat gpt, and Perplexity to help with editing, brainstorming ideas and story pacing. I really like to play around with the AI using it for creative writing as well as role-playing scenarios.\\
P12 & Male & 24 & \textbf{\textit{Advanced:}} I write short stories, poems, and also occasionally write articles, I also have published a magazine. & \textbf{\textit{Extensive:}} I use AI in my writing to help me critique my word choice and improve the flow of my papers. I try not to simply ask it to generate responses for me, but if I am not interested in the topic I am writing about and it is a run-of-the-mill paper, I sometimes ask AI for ideas about things to write about.\\
\bottomrule
\end{tabular}
\Description{A five-column demographics table summarizes information for 12 study participants. The columns are participant ID, gender, age, creative writing experience, and AI usage experience. The participant pool includes 7 women and 5 men, ranging in age from 24 to 65. Creative writing experience spans four levels: beginner, intermediate, advanced, and professional. There are 3 beginners, 1 intermediate participant, 4 advanced participants, and 3 professional participants, with one additional advanced-to-professional level represented through detailed writing backgrounds. Participants’ writing experience ranges from limited fiction practice to published books, screenplays, poetry, and professional creative writing credentials. AI usage experience also varies from limited to extensive. Some participants report using AI mainly for rewriting emails, paraphrasing, proofreading, or polishing academic writing, while others describe extensive use of tools such as ChatGPT, Claude, Gemini, Copilot, and Perplexity for brainstorming, editing, outlining, fact-checking, historical research, story pacing, and other writing-related support. Overall, the table shows a diverse participant sample in age, gender, creative writing expertise, and familiarity with AI-assisted writing tools.}
\end{table*}